\documentclass[sigconf]{acmart}
\usepackage{enumitem}
\usepackage{subcaption}
\usepackage{adjustbox}
\AtBeginDocument{}

\setcopyright{none}
\renewcommand{\footnotetextcopyrightpermission}[1]{}
\copyrightyear{2018}
\acmYear{2018}
\acmDOI{XXXXXXX.XXXXXXX}
\acmConference[Conference acronym 'XX]{Make sure to enter the correct
  conference title from your rights confirmation email}{June 03--05,
  2018}{Woodstock, NY}
\acmISBN{978-1-4503-XXXX-X/2018/06}

\colorlet{blue}{black}
\begin{document}

\title{Beyond Hotness: High-Resolution Memory Characterization of Cloud and LLM Workloads for Tiered Memory with ARM SPE}

\author{Jie Ren}
\affiliation{%
  \institution{William \& Mary}
  \country{USA}
}
\email{jren03@wm.edu}

\author{Ivy Peng}
\affiliation{%
  \institution{KTH}
  \country{Sweden}
}
\email{ivybopeng@kth.se}

\begin{abstract}
Memory demands from data analytics and LLM workloads grow rapidly, driving up memory costs in cloud and data centers. Multi-tier memory has emerged as a promising solution to balance performance and cost. However, effectively leveraging it requires a deep understanding of workload memory behaviors. This work introduces a high-resolution memory characterization methodology for analyzing workload-specific opportunities for multi-tier memory usage. In response to the increasing deployment of ARM-based processors in data centers, we implement our methodology using the recent Statistical Profiling Extension (SPE) for ARM. By characterizing eight memory-intensive workloads from CloudSuite and LLM applications, we identify region-specific access patterns, evaluate the effectiveness of cloud-oriented cache hierarchies, and propose new metrics based on memory idle periods and latency profiles to determine suitable data placement for tiered memory.
\end{abstract}


\maketitle

\section{Introduction}
Memory-intensive workloads are becoming increasingly important to cloud and data centers (DCs). The rapid proliferation of large language models (LLMs), alongside traditional workloads such as real-time analytics, in-memory databases, and graph processing, continues to drive up the demand for high-capacity, high-performance memory. At the same time, the rising procurement and operational costs of memory have made multi-tier memory system an attractive solution for meeting these demands cost-effectively~\cite{peng2018siena,peng2018characterizing,xie2021md}. Multi-tier memory systems pair a fast but expensive memory tier with a larger, more affordable, but slower tier\textcolor{blue}{, striking a balance} between cost and performance~\textcolor{blue}{\cite{atc21_autotiering, sosp21_hemem, asplos17_thermostat, eurosys24-mtm}}. A notable example is the integration of high-bandwidth memory (HBM) with DDR memory in commercial systems such as Intel’s Sapphire Rapids (Xeon Max Series), which combines HBM with DDR5. Another emerging approach is disaggregated memory, enabled by Compute Express Link (CXL), which offers high-bandwidth, low-latency cache-coherent interconnects to implement memory pooling at rack-scale~\cite{maruf2023tpp, wahlgren2023quantitative, pond}.  

Effectively managing workloads in multi-tier memory systems requires a deep understanding of their memory access patterns and the interplay between hardware and system software. OS kernel and hardware performance counters are two main mechanisms for building tools for memory profiling and characterization. For instance, Linux's DAMON~\cite{park2019profiling,palit16} (Data Access MONitor) is a kernel feature that can monitor memory access patterns within a given overhead budget by trading off accuracy for performance. Many profiling tools are based on linear scanning of PTEs (Page Table Entries) to reveal page-level access and residency information. When Performance Monitoring Units (PMUs) are augmented with information, such as virtual addresses, latency, and program counters, as presented by Processor Event-Based Sampling (PEBS) on Intel processors or Instruction-Based Sampling (IBS) on AMD processors, they can provide detailed low-level information.

Recently, data centers (DCs) and cloud providers have increasingly adopted high-end ARM processors as an alternative to x86 platforms~\cite{kobayashi2026performance}. \textcolor{blue}{This trend includes ARMv8.x processors targeting cloud and compute-intensive workloads.} For instance, Fugaku supercomputer is built on ARM A64FX processors. NVIDIA and Amazon have also used ARM for their server-class processors, including NVIDIA Grace (used in the Jupiter supercomputer) and AWS Graviton. Despite this growing momentum, most existing memory profiling and analysis tools are specific to x86 architectures. As a result, ARM-based platforms have remained relatively underexplored in terms of fine-grained memory characterization, partly due to the limited support of precise sampling features on earlier ARM implementations. With the introduction of the Statistical Profiling Extension (SPE) in ARMv8.2, detailed memory analysis on ARM platforms becomes feasible. 

\begin{figure*}
    \centering
    \includegraphics[width=\linewidth]{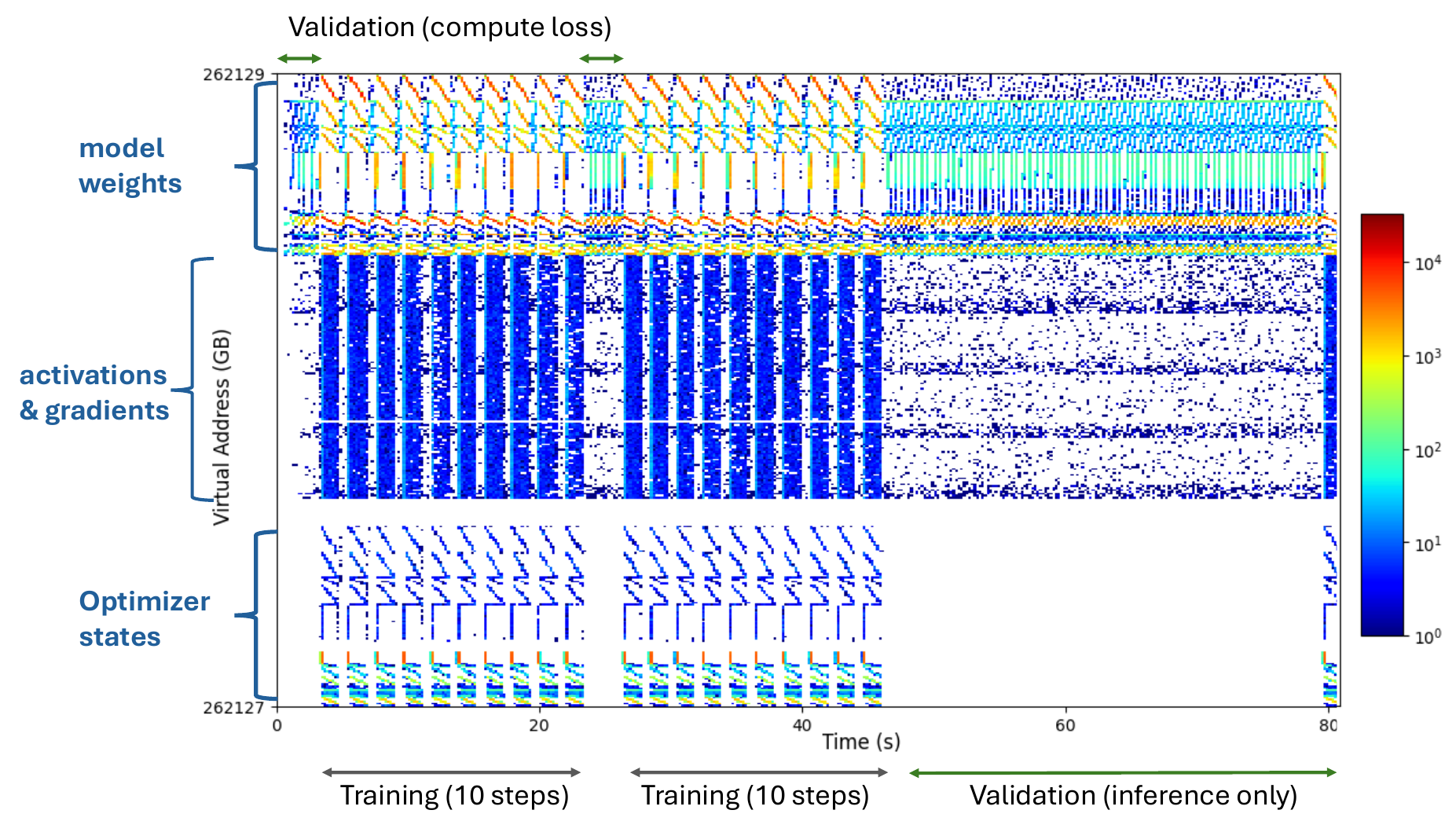}
    \caption{\textcolor{blue}{L2 cache hits by virtual address over time during GPT2 training. Forward-only intervals include validation-loss computation and text generation.}}
    \label{fig:l2-llm}
\end{figure*}

This work fills the gap by providing a memory characterization method on ARM-based systems that incorporate SPE. First, we \textcolor{blue}{propose} a methodology for understanding memory-intensive workloads from two complementary perspectives\textcolor{blue}{: workload characteristics and memory hierarchy utilization.} Second, we then characterize eight memory-intensive workloads using a set of metrics tied to efficient utilization of multi-tier memory. Finally, we showcase integrating the SPE profiling analysis into a data migration optimization runtime. Figure~\ref{fig:l2-llm} presents an example of detailed memory access analysis in LLM training, showing L2 cache hits associated to virtual memory address over time, \textcolor{blue}{distinguishing three groups of memory regions (weights, activations/gradients, and optimizer states) and their distinct access patterns across training and forward-only phases.} 

In summary, we made the following contributions in this work.

\begin{itemize}[leftmargin=*]
    \item We provide a high-resolution memory characterization method comprising a profiling tool and comprehensive analysis framework targeting ARM processors.
    \item We provide a high-resolution profiling method based on ARM SPE, and quantify its accuracy and overhead.     
    \item We characterize eight memory-intensive workloads from Cloud and LLMs on a data center-grade ARM-based platform.    
    \item We propose a set of workload-focused metrics for assessing the feasibility of multi-tier memory and \textcolor{blue}{evaluate their usefulness through experiments with an SPE-guided data migration runtime.}
\end{itemize}

\section{Characterization Methodology}

Our memory characterization serves two objectives. The first aims to provide insights for deriving effective prefetching strategies to optimize workloads on multi-tier memory. For this, access patterns inherent in workload characteristics need to be identified comprehensively. The second aims to provide insights for optimizing and tailoring cache and memory configurations in future system designs for specific workloads. For this, accurately monitoring utilization and identifying hotspots at each level of the cache and memory hierarchy is desirable. To support these goals on real-world workloads, we define the following design principles. 

\begin{enumerate}[leftmargin=*]
    \item \textcolor{blue}{\textbf{Comprehensive Characteristics.}} Capture all workload access patterns critical for deriving prefetching and data placement strategies, including execution phases, temporal locality, spatial locality, and hot and cold regions.
    \item \textcolor{blue}{\textbf{Memory Hierarchy Awareness.}} Quantify the utilization of the complex cache hierarchy, coherence interconnects, and main memory, and correlate \textcolor{blue}{this utilization with} specific memory regions.
    \item \textcolor{blue}{\textbf{Support High Concurrency.}} Support isolating per-thread memory access information in highly concurrent workloads using tens and hundreds of threads in data centers and cloud environments.
    \item \textcolor{blue}{\textbf{High Resolution.}} Support high resolution in time to detect fast phase changes in real applications, and high resolution in memory space to reduce fragmentation.
    \item \textcolor{blue}{\textbf{Low Overhead.} Incur low resource and runtime overhead during workload execution.}
    \item \textcolor{blue}{\textbf{Programming Language Agnostic.}} Support diverse cloud workloads in different programming languages and frameworks transparently.
\end{enumerate}

\subsection{An ARM SPE-Based Design and Implementation}
Guided by these design principles, we compare various memory profiling mechanisms, including \textit{binary instrumentation}, \textit{compiler analysis}, \textit{OS kernel}, and \textit{Performance Monitoring Unit (PMU)}. Binary instrumentation and compiler analysis offer detailed information at the memory-instruction level, but lack visibility into architectural information of the cache and memory hierarchy, %
thus violating principle (2). OS kernel-based methods, such as DAMON~\cite{park2019profiling}, leverage page-table entry (PTE) scans for region-based profiling, identifying only page-level access patterns of frequently accessed regions. This approach results in coarse-grained profiling and misses detailed insights into less active or quickly changing regions, violating principles (1)--(4). Given these limitations, we adopt a PMU-based mechanism leveraging the ARM SPE extension to develop our memory characterization method.

ARM SPE is a recent extension to the PMU that augments hardware-assisted CPU profiling with detailed information, such as virtual memory addresses, latency, and events throughout the execution pipeline. It was introduced in the Armv8-A architecture and first became available with the Neoverse N1 CPU. As an instruction sampling-based mechanism, ARM SPE selects sampled instructions at a user-defined interval, i.e. the sample period $N$. \textcolor{blue}{The sampling counter selects an operation approximately every $N$ eligible operations, with randomization to reduce bias, and tracks it through the execution pipeline.} If it is a memory instruction, \textcolor{blue}{ARM SPE can capture detailed information such as supported cache and TLB events, a data-source identifier, and latency counts for the sampled operation.} \textcolor{blue}{A sample record is created and selectively filtered before saving it into a buffer in memory as a sequence of \textit{packets}.} Filtering options relevant to memory characterization include events such as L1 Data (L1D) cache misses and Translation Lookaside Buffer (TLB) misses. The main tradeoff between accuracy and overhead is determined by the sampling period $N$, which has a machine-specific minimum value. %

We designed and implemented the SPE-based memory profiling tool to support transparent profiling by preloading the library using the dynamic linker. Environment variables are used to adapt the profiling configuration, e.g. the sample period. A key difference between Intel's PEBS and ARM SPE is that PEBS is event-based while SPE is instruction-based. \textcolor{blue}{PEBS triggers records from selected performance events, while SPE selects operations for profiling. An SPE sample record can combine address, event, data-source, and latency packets for the same operation~\cite{arm_spe_analysis}.}

\textcolor{blue}{Our tool separates samples on different cores, supporting per-thread analysis when worker threads are pinned to cores.} With this capability, we can detect workload imbalance among threads and optionally trace only a subset of the cores to reduce overhead in symmetric workloads. Internally, we implement the tool with separate staging buffers to avoid synchronization among threads for high scalability. The characterization supports high resolution in time, with profiling windows as fine-grained as 200 ms. The resolution of memory address space is at the byte-addressable level. The tool reduces performance overhead by carefully selecting event filters, sample frequency, and buffer size. For memory characterization in this work, we set the SPE filter to sample L1D load misses \textcolor{blue}{with a sampling period of 1,024} and use a 1~MiB buffer size. %

We leverage the sampled data source information to correlate the utilization of cache and memory hierarchy to specific memory regions. In particular, the supported hierarchical information includes L1D, L2, peer CPU, local cluster, system cache, peer cluster, and DRAM. %

We further study DRAM accesses as they incur the longest latency, by capturing \textcolor{blue}{three latency counts:} issue latency, address translation latency, and total latency. Issue latency measures the number of cycles waiting for the instruction to be issued. Total latency measures the number of cycles to execute the instruction. \textcolor{blue}{For samples attributed to DRAM, the reported latency is an SPE operation-latency measurement and can include delays beyond DRAM service time.} With this detailed latency information, we can identify problematic memory regions that suffer from high latency.

\subsection{Validation}
\label{sec:method}

\textcolor{blue}{In this section, we thoroughly evaluate our proposed memory characterization method,
providing a validation of ARM SPE’s suitability for optimizing multi-tier memory on ARM-based platforms.}

To validate correctness, we use three benchmarks with well-known access patterns as ground truth. \textit{STREAM} is a widely-used benchmark that employs sequential access patterns for measuring peak memory bandwidth. \textcolor{blue}{We use its multi-threaded version to validate our tool's ability to accurately capture per-thread streaming accesses.} \textcolor{blue}{\textit{RandomAccess} performs random access to a large heap-allocated array following a normal distribution, which is used to validate our tool's efficiency in capturing random access patterns with varying spatial localities.} \textcolor{blue}{Additionally, \textit{MASIM}~\cite{nair2024telescope} provides a phase-based benchmark that emulates real-world workload behavior through configurable hot regions and access phase transitions, enabling validation of dynamic memory access pattern tracking.} 
We quantify our tool's accuracy using precision and recall for hot page identification, and measure runtime overhead across various sampling rates. We compare both accuracy and overhead against DAMON~\cite{park2019profiling}, a state-of-the-art OS kernel-based profiling tool.

\begin{figure}[!htbp]
    \centering
    \includegraphics[width=\linewidth]{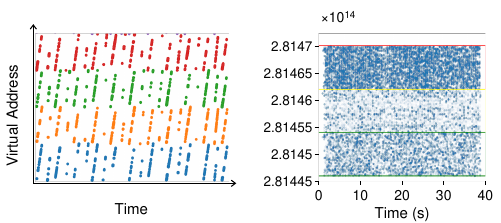}
    \caption{\textcolor{blue}{The scatter plot of L1D misses in STREAM, exhibiting temporal and spatial locality. Each color represents a different thread.}}
    \label{fig:stream}
\end{figure}

\textbf{Access patterns.} We assess whether the tool can effectively capture workload access patterns in terms of spatial and temporal locality and execution phases. 
Figure~\ref{fig:stream} presents the memory access patterns captured in STREAM. The left plot shows the distribution of accessed virtual addresses (y-axis) over execution time (x-axis), with each color representing a different thread. The visualization reveals how each thread accesses a contiguous memory region in a streaming pattern, with execution phases repeating periodically. Notably, such fine-grained per-thread access patterns cannot be detected by OS kernel-level mechanisms such as DAMON. The right plot displays access density across the virtual address space over time, showing that array c (higher addresses) receives more accesses than arrays a and b (lower addresses), \textcolor{blue}{reflecting different sampled access counts among the arrays in this run.}

\begin{figure}[!htbp]
    \centering
    \includegraphics[width=\linewidth]{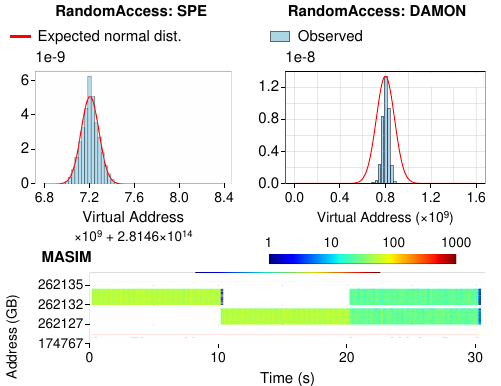}
    \caption{\textcolor{blue}{Top two: memory access histograms from a SPE-based tool and DAMON on RandomAccess. Bottom: memory access heatmap from the SPE-based tool on MASIM, showing three access phases across different address ranges.}}
    \label{fig:distribution}
\end{figure}

Figure~\ref{fig:distribution} compares the access distributions captured by our SPE-based tool (\textcolor{blue}{top left}) and DAMON (\textcolor{blue}{top right}) for the RandomAccess benchmark. The left plot shows that the ARM SPE tool accurately reconstructs the normal distribution of memory accesses, with the captured samples (blue bars) closely matching the theoretical distribution (red line). This demonstrates that ARM SPE can reliably capture different hotness levels across the memory space, which is critical for page placement decisions in multi-tier memory systems. In contrast, DAMON's captured distribution (\textcolor{blue}{top right plot}) shows significant information loss, particularly for pages with moderate to low access frequencies. This limitation is due to DAMON's PTE-scanning approach, which has lower sampling coverage and easily misses infrequent accesses during its periodic scans.


Figure~\ref{fig:distribution} (\textcolor{blue}{bottom}) shows the memory access heatmap captured by our SPE-based tool for MASIM. The heatmap clearly reveals distinct access patterns across three execution phases: the first phase exclusively accesses array a, the second phase only accesses array b, and the third phase accesses both arrays equally. These phase transitions are captured with clear boundaries, demonstrating the ability to track dynamic changes in memory access patterns over time with our tool. \textcolor{blue}{Under the tested configuration, DAMON does not clearly resolve these phase changes.}

\begin{figure}[!htbp]
    \centering
    \includegraphics[width=\linewidth]{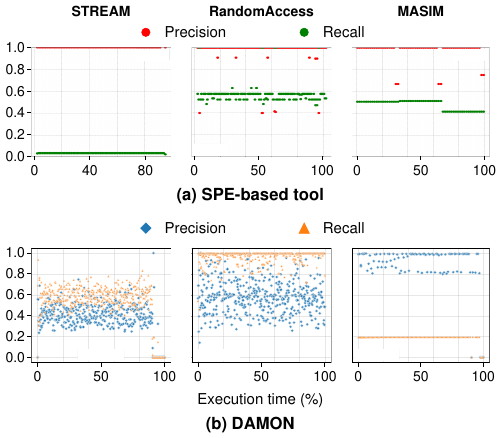}
    \caption{\textcolor{blue}{Precision and Recall measured over time in three benchmarks. Top row: SPE-based tool; bottom row: DAMON. SPE-based tool achieves perfect precision with stable recall.}}
    \label{fig:precision_recall}
\end{figure}

\textbf{Accuracy. }We use the \textit{Precision} and \textit{Recall} metrics to quantify the accuracy of identifying hot regions in workloads. Precision is calculated in a profiling window as the fraction of identified hot memory regions that are truly hot, while recall measures the fraction of actual hot regions that are correctly identified. We establish ground truth based on application semantics derived from source code analysis. In STREAM, the memory regions corresponding to the three heap-allocated arrays are considered hot as they are continuously accessed throughout execution. In RandomAccess, we define memory regions within one standard deviation (1$\sigma$) of the mean in the normal distribution as hot, capturing 68.3\% of all accesses. MASIM's ground truth varies in the three phases. 

Figure~\ref{fig:precision_recall} presents the precision and recall measurements over time for both our SPE-based tool (top row) and DAMON (bottom row) across the three benchmarks.
For our SPE-based tool, precision consistently maintains 1.0 throughout execution in all three benchmarks, indicating that all identified hot memory regions are truly hot. This ensures that page migration decisions would not mistakenly move cold regions into the fast tier. The recall achieves 0.5 and 0.6 for RandomAccess and MASIM respectively, but only 0.02 for STREAM. This variation occurs because we classify the top 50 sampled regions as hot at each time step, which limits the number of selected hot regions and causes low recall in benchmarks with many uniformly accessed memory regions like STREAM. DAMON achieves comparable recall for most benchmarks due to its region-based profiling that naturally covers larger memory areas. However, it exhibits low precision and high variability in both metrics within the same workload. This instability, particularly shown in the scattered point distributions, demonstrates DAMON's difficulty in consistently identifying hot regions across different access patterns.

\begin{figure}[!htbp]
    \centering
    \includegraphics[width=0.9\linewidth]{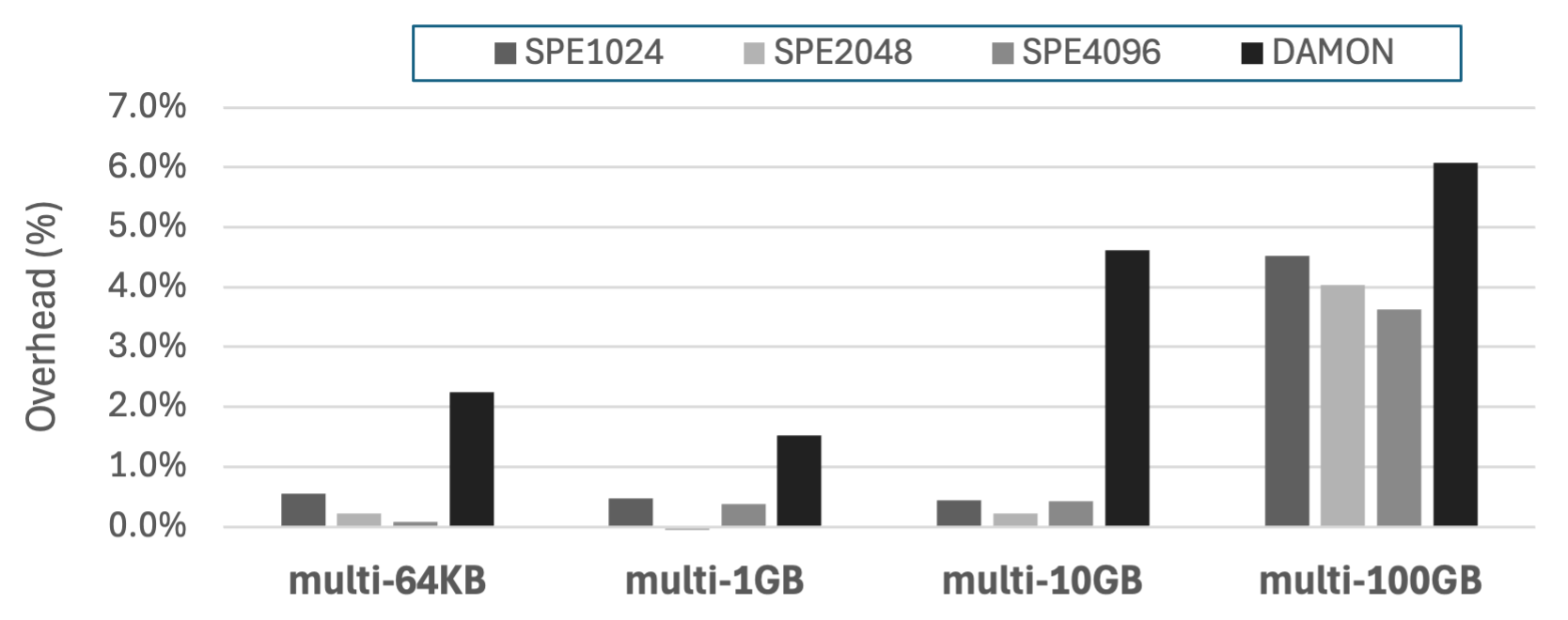}
    \caption{\textcolor{blue}{The performance overhead of ARM SPE using three sample periods and DAMON in four MASIM configurations.}}
    \label{fig:overhead}
\end{figure}

\textbf{Overhead.} We evaluate performance overhead using MASIM configured with four problem sizes ranging from 64~KB to 100~GB regions. We use three different sampling periods for our SPE-based profiling tool and DAMON. \textcolor{blue}{The results in Figure~\ref{fig:overhead} show that our SPE-based approach incurs less than 0.6\% overhead in the 64~KB, 1~GB, and 10~GB configurations, while the 100~GB configuration reaches 3.6--4.5\%.} Overhead remains consistent across different SPE sampling intervals (1024, 2048, 4096) in 64~KB, 1~GB and 10~GB MASIM, while overhead decreases with larger sampling intervals in 100~GB MASIM with SPE4096 showing the lowest overhead and SPE1024 the highest. \textcolor{blue}{Under the tested configurations, DAMON incurs about 1.5--6.1\% runtime overhead.}

\section{Experimental Methodology}
\label{sec:setup}

We use a testbed based on ARM Ampere® Altra® Max 64-Bit~\cite{ampere} processor with 128 \textcolor{blue}{Neoverse N1 (Armv8.2+) cores} running at 3.0~GHz. Every two cores are co-located within a DSU (DynamiQ Shared Unit). The cache hierarchy consists of 64~KB L1I and 64~KB L1D per core, 1~MB L2 per core, and a shared 16~MB System Level Cache (SLC) as the last-level cache (LLC). The main memory has a total of 512~GB DDR4 and 200~GB/s peak bandwidth. This cache hierarchy configuration, in contrast to shared L2 and large LLC often used in HPC systems, is tailored for multi-tenant environments in data centers. The testbed runs Ubuntu 22.04 with Linux kernel 5.15. Applications were built using GCC 11.4, OpenMPI v4.1.2, and OpenBLAS 0.3.20. We also evaluated on an AmpereOne® processors to ensure portability and traces are similar to those obtained on Ampere® Altra® testbed and hence omitted.%
We evaluate eight workloads commonly deployed in data center environments using our SPE-based profiling tool, covering LLM training and inference, graph analytics, and data processing applications, as summarized in Table~\ref{tab:apps}.

\begin{table}[!htbp]
\centering
\caption{\textcolor{blue}{Memory-intensive applications for characterization.}} 
\begin{adjustbox}{max width=\linewidth}
\begin{tabular}{lll}
\toprule
\textbf{Application} &\textbf{Language} &\textbf{Domain}\\\midrule%
\textcolor{blue}{Llama 8B}      &\textcolor{blue}{C++}    &LLM Inference \\
GPT2         & C          &LLM Pretraining \\%
PageRank (PR)            &Scala, Java       &Graph Analytics\\%
Connected Components (CC) &Scala, Java  &Graph Analytics \\%
Triangle Counting (TC)       &Scala, Java  &Graph Analytics\\%
In-Memory Analytics               &Java        &Data Processing \\%
Memcached       &C              &Data Caching \\%
Cassandra       &Java    &Data Serving \\%
\bottomrule
\end{tabular}
\end{adjustbox}
\label{tab:apps}
\end{table}

\begin{figure*}[!htbp]
    \centering
    \begin{subfigure}{0.24\textwidth}
        \includegraphics[width=\textwidth]{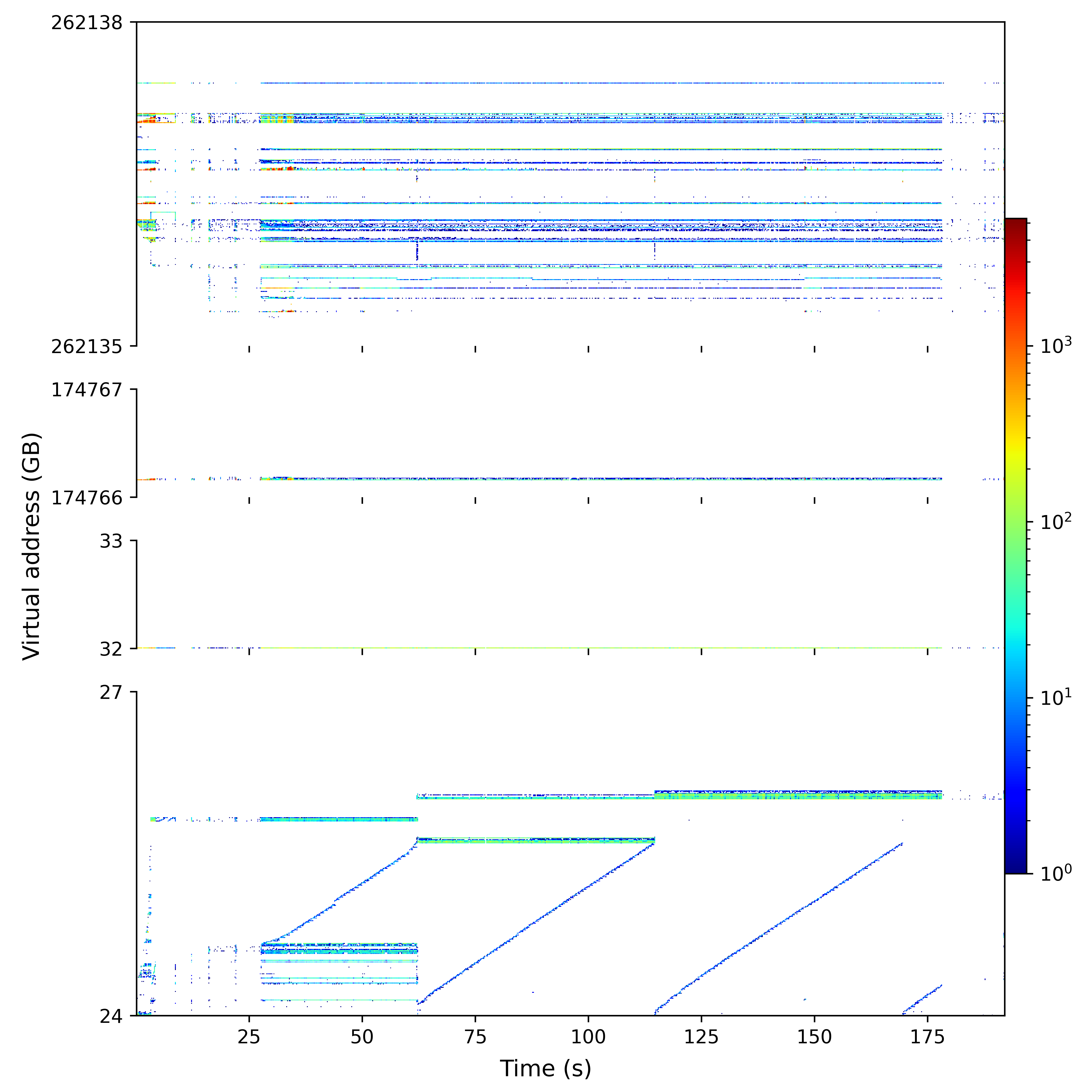}
    \caption{Cassandra}
    \label{fig:cassandra}
    \end{subfigure}
    \begin{subfigure}{0.24\textwidth}
        \includegraphics[width=\textwidth]{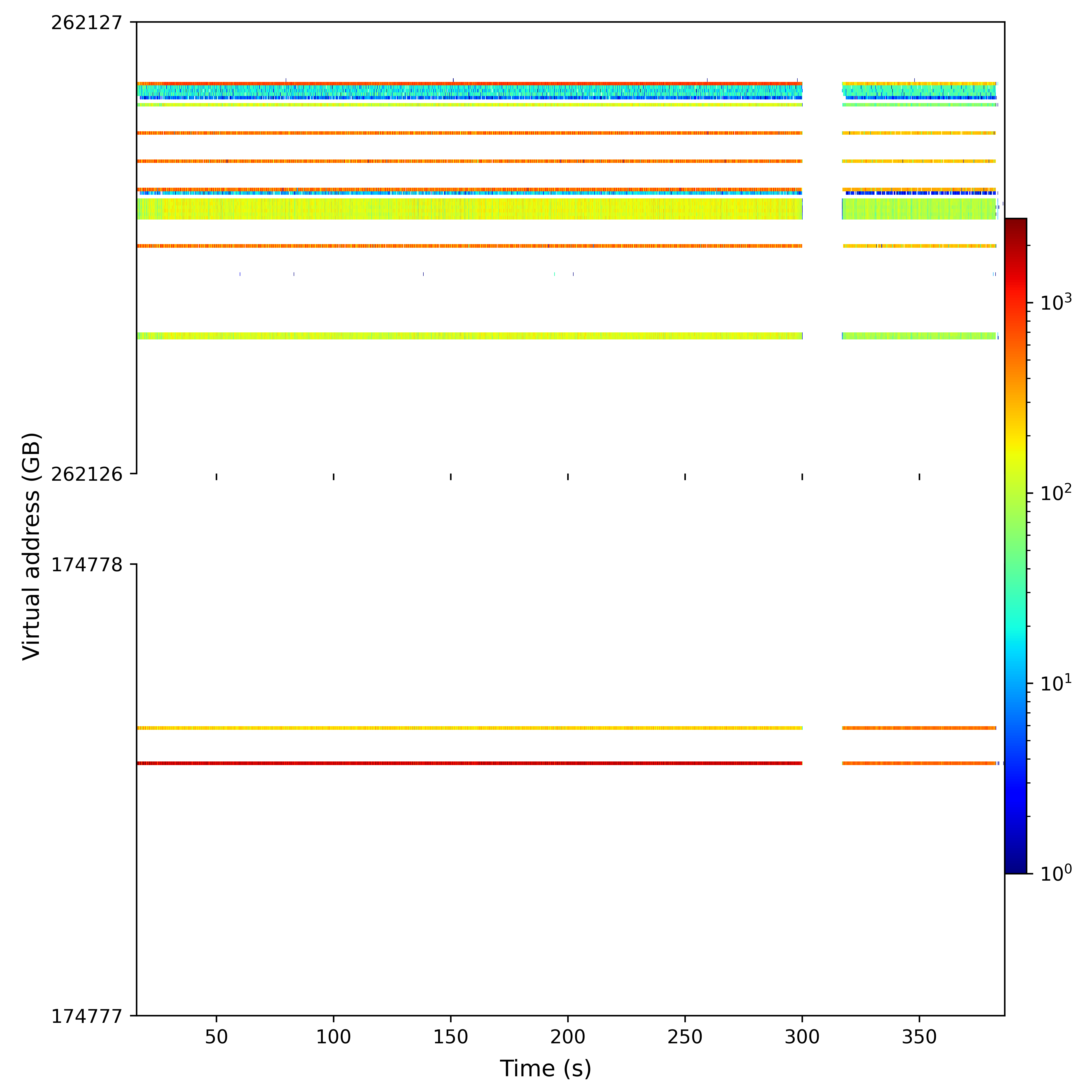}
    \caption{Memcached}
    \label{fig:memcached}
    \end{subfigure}    
    \begin{subfigure}{0.24\textwidth}
        \includegraphics[width=\textwidth]{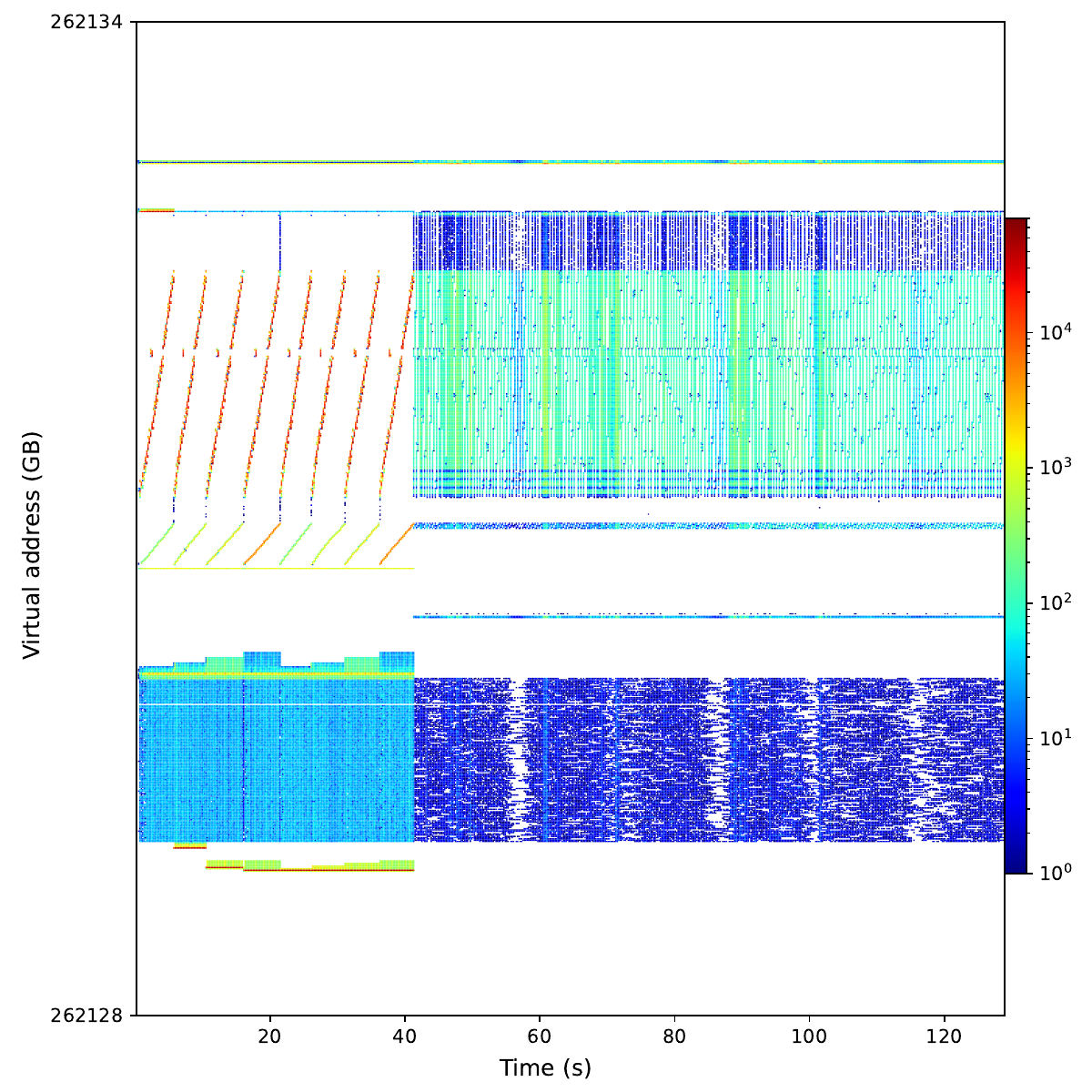}%
        
    \caption{\textcolor{blue}{Llama 8B Inference}}
    \label{fig:llama}
    \end{subfigure}
    \begin{subfigure}{0.24\textwidth}
        \includegraphics[width=\textwidth]{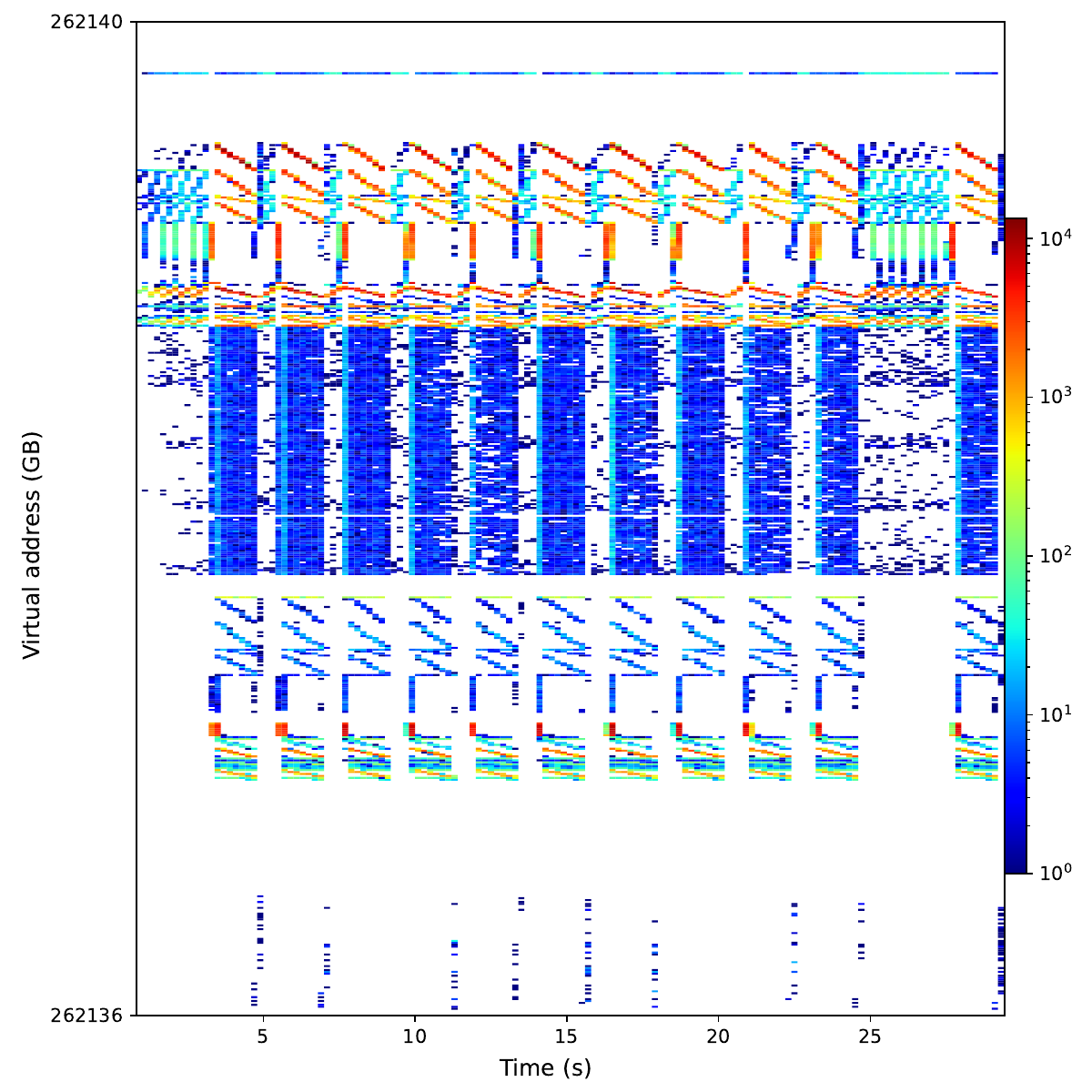}%
    \caption{GPT2 Pretraining}
    \label{fig:gpt2}
    \end{subfigure}    
    \begin{subfigure}{0.24\textwidth}
        \includegraphics[width=\textwidth]{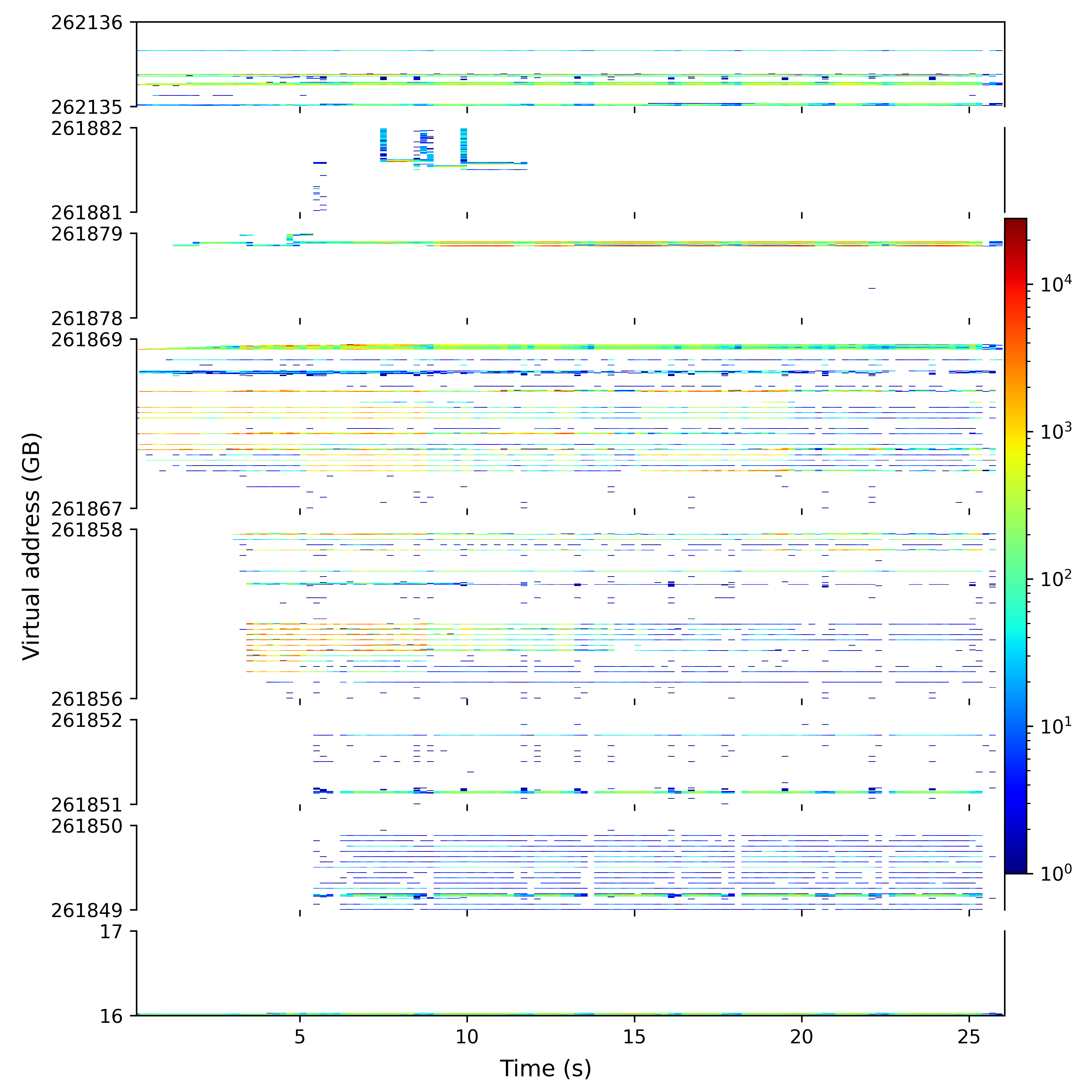}
    \caption{In-Memory Analytics}
    \label{fig:analytics}
    \end{subfigure}
    \centering
    \begin{subfigure}{0.24\textwidth}
        \includegraphics[width=\textwidth]{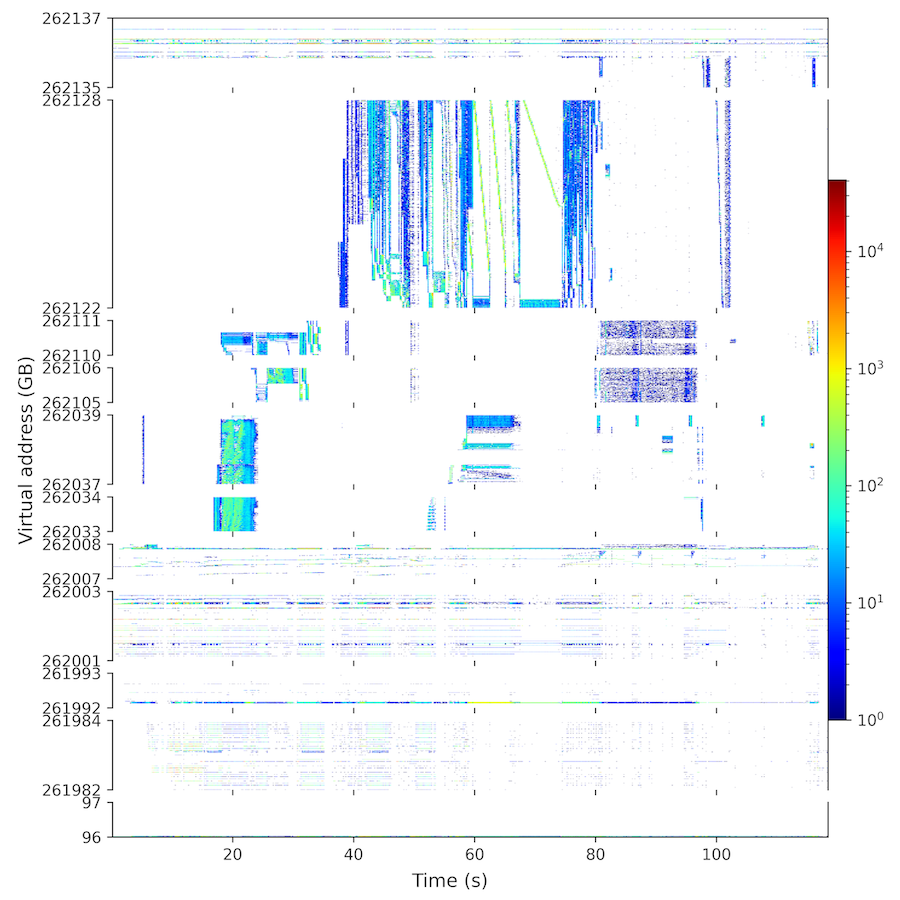}
    \caption{Triangle Counting}
    \label{fig:tc}
    \end{subfigure}
    \begin{subfigure}{0.24\textwidth}
        \includegraphics[width=\textwidth]{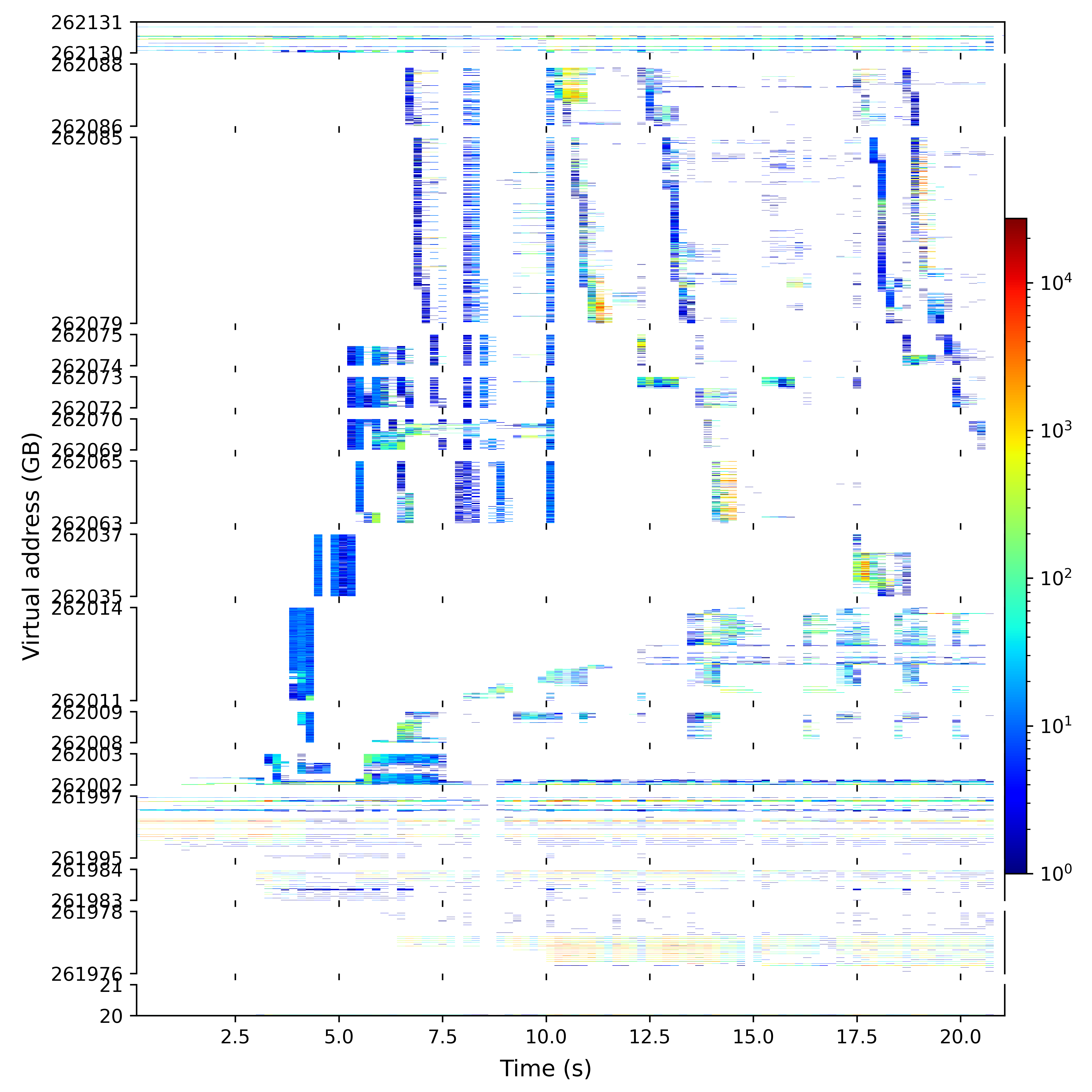}
    \caption{Page Rank}
    \label{fig:pr}
    \end{subfigure}
    \begin{subfigure}{0.24\textwidth}
        \includegraphics[width=\textwidth]{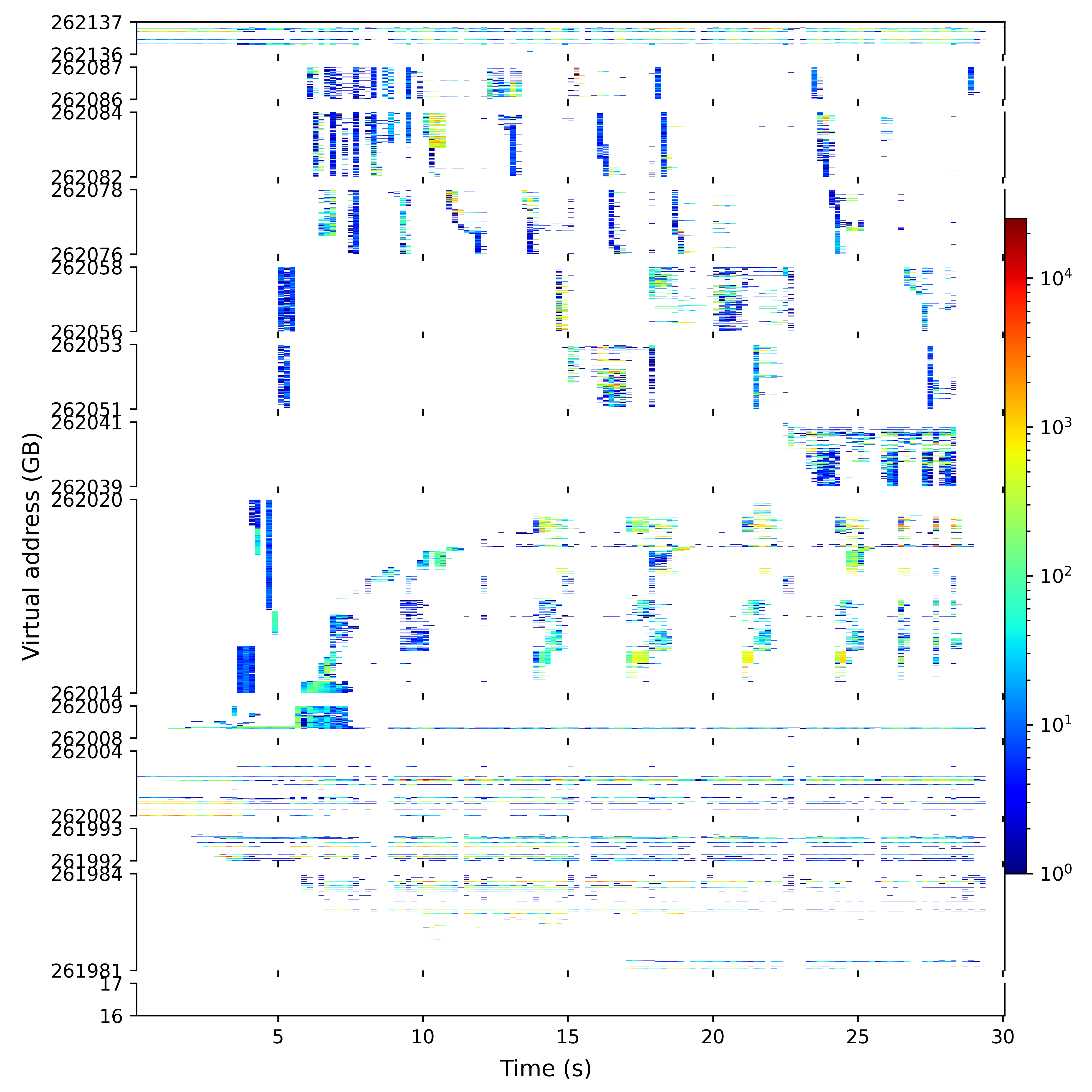}
    \caption{Connected Components}
    \label{fig:cc}
    \end{subfigure}    
    \caption{\textcolor{blue}{Sampled L1D load misses by virtual address (sampling period: 1024; bins: 200~ms by 8~MiB).}}%
    \label{fig:pattern}
\end{figure*}

\textbf{LLM Workloads}. For inference, we use \textcolor{blue}{Llama 8B model} (denoted as Llama) with the llama.cpp~\cite{llama_cpp} CPU implementation. \textcolor{blue}{We use quantized model weights while keeping the KV cache in float16} to ensure KV cache access patterns remain visible in the memory footprint. We evaluate with batch size 8, processing 128 prompt tokens to generate 100 tokens on 16 threads, measuring performance in tokens per second. For training, we evaluate GPT-2 124M pretraining using the \texttt{train\_gpt2} in llm.c~\cite{llm.c} CPU implementation, configured with maximum sequence length of 1024 and \textcolor{blue}{1192 batches in the training dataset}, reporting execution time. 

\textcolor{blue}{To further examine KV-cache access patterns during inference (Figure~\ref{fig:kvcache}), we use Qwen3-30B-A3B-Instruct-2507~\cite{qwen3_2507} with the llama.cpp CPU implementation on the same testbed and with the same SPE sampling period of 1,024. This mixture-of-experts model uses UD-IQ2\_XXS GGUF weights (2.71 effective bits per weight) and FP16 K and V caches. We process 2,000 prompt tokens and generate 512 tokens with one sequence on 128 threads. The context capacity is 4,096 tokens, with a token batch limit of 2,048 and a microbatch limit of 512.}

\textbf{Graph Analytics.} We evaluate three fundamental graph algorithms: PageRank for computing node importance in web ranking applications, Triangle Counting for community detection and network analysis, and Connected Components for identifying clusters within graphs. All three workloads process a Twitter dataset using 32 threads and 128 GB memory.

\textbf{Data Processing.} We evaluate three widely used data processing applications. Memcached runs as an in-memory key-value store with a Twitter dataset scaled 112×, using approximately 40 GB memory. After warmup, it operates with a 0.8 get-set ratio, reporting 99th percentile latency. Cassandra, evaluated using YCSB, processes 100K records with 8 threads after warmup, measuring throughput in operations per second. For in-memory analytics, we run Apache Spark MLlib's movie recommendation system using the MovieLens dataset, configured with 32 threads and 256 GB memory allocation, measuring execution time for generating top-50 recommendations.

\section{Memory-intensive Workloads}
We apply the characterization methodology to eight memory-intensive workloads. Our characterization study aims to provide insights and metrics for workloads on multi-tier memory systems, such as HBM-DRAM and CXL-based disaggregated memory systems. Therefore, we focus on three aspects of the characterization: workload-level characteristics, memory hierarchy utilization, and memory region profiles. 

\subsection{Workload Insights}
\label{sec:pattern}
\textcolor{blue}{We examine memory access patterns based on temporal and spatial locality characteristics, including strided, sparse, and block patterns.} Figure~\ref{fig:pattern} presents memory access \textcolor{blue}{sample-count heatmaps filtered for L1D load misses.}

\textbf{Pattern I: Strided Pattern} represents memory accesses with consistent address offsets. In the heatmaps, these appear as horizontal lines (i.e., stride-0, indicating repeated access to the same address over time) or diagonal lines (i.e., \textcolor{blue}{stride-$n$}, indicating traversal with fixed address increments).

\begin{figure}[!htbp]
    \centering
    \includegraphics[width=0.78\linewidth]{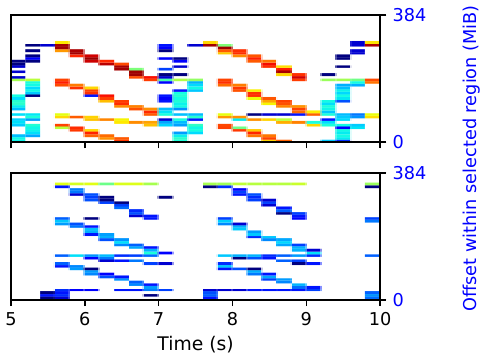}
    \vspace{-5pt}
    \caption{\textcolor{blue}{GPT-2 pretraining accesses at 5--10~s. Upper: model parameters; lower: optimizer states. Each panel spans 384~MiB with its own address origin.}}
    \vspace{-10pt}
    \label{fig:stride}
\end{figure}

As an example of strided patterns, we examine high-resolution profiling results from our SPE-based profiling tool. Figure~\ref{fig:stride} zooms into specific memory regions during GPT-2 pretraining. The upper region exhibits V-shaped diagonal patterns from model parameter accesses: ascending diagonals during forward passes and descending diagonals during backward passes, \textcolor{blue}{reflecting the opposite layer traversal order of forward and backward passes.} The lower region shows optimizer state accesses \textcolor{blue}{after the backward pass}. \textcolor{blue}{After} the backward pass computes gradients layer by layer (from last to first), the optimizer uses these gradients to \textcolor{blue}{update its internal states}. The cooler colors in the lower region indicate \textcolor{blue}{fewer sampled memory accesses per bin compared to the upper region}.

Similar strided patterns appear across diverse workloads. Cassandra (Figure~\ref{fig:cassandra}) exhibits horizontal strided patterns in upper address ranges from repeated metadata structure accesses, while diagonal patterns in lower addresses indicate systematic memory traversal during compaction operations. Memcached (Figure~\ref{fig:memcached}) shows prominent horizontal strided patterns with highly concentrated hot bands, indicating specific hot keys and critical metadata structures being repeatedly accessed. In-Memory Analytics (Figure~\ref{fig:analytics}) displays numerous horizontal strided patterns distributed across the address space with varying intensities, reflecting access to multiple data structures with different usage frequencies. Graph analytics workloads (Figures~\ref{fig:tc}, \ref{fig:pr}, and~\ref{fig:cc}) also display horizontal strided patterns in their lower address spaces.

\textbf{Pattern II: Sparse Pattern} represents memory accesses to small, irregular subsets of addresses within a region over time. In the heatmaps, these appear as scattered dots or isolated points with no clear continuity in either temporal (horizontal) or spatial (vertical) dimensions, indicating infrequent and non-contiguous memory usage.

\begin{figure}[!htbp]
    \centering
    \includegraphics[width=0.78\linewidth]{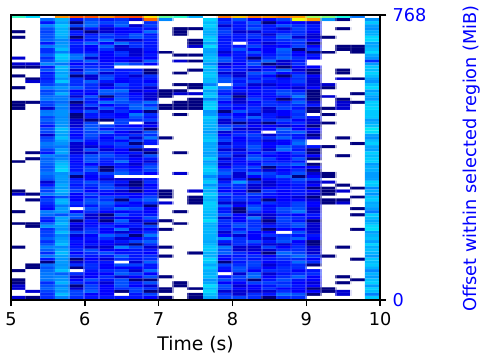}
    \caption{\textcolor{blue}{GPT-2 activation/gradient accesses at 5--10~s over 768~MiB: sparse patterns during forward passes and block patterns during backward passes.}}
    \vspace{-10pt}
    \label{fig:sparse}
\end{figure}

To illustrate sparse patterns, Figure~\ref{fig:sparse} zooms into the activation/gradient memory region during GPT-2 pretraining. During forward passes, activations (visible as scattered blue dots) \textcolor{blue}{exhibit sparse sampled access patterns in these intervals.} Similar sparse patterns emerge during validation phases as shown in Figure~\ref{fig:gpt2}, which consist of forward computation only, and in other scenarios where data structures are accessed infrequently or irregularly.

\begin{figure*}[!htbp]
    \centering
    \includegraphics[width=0.9\textwidth]{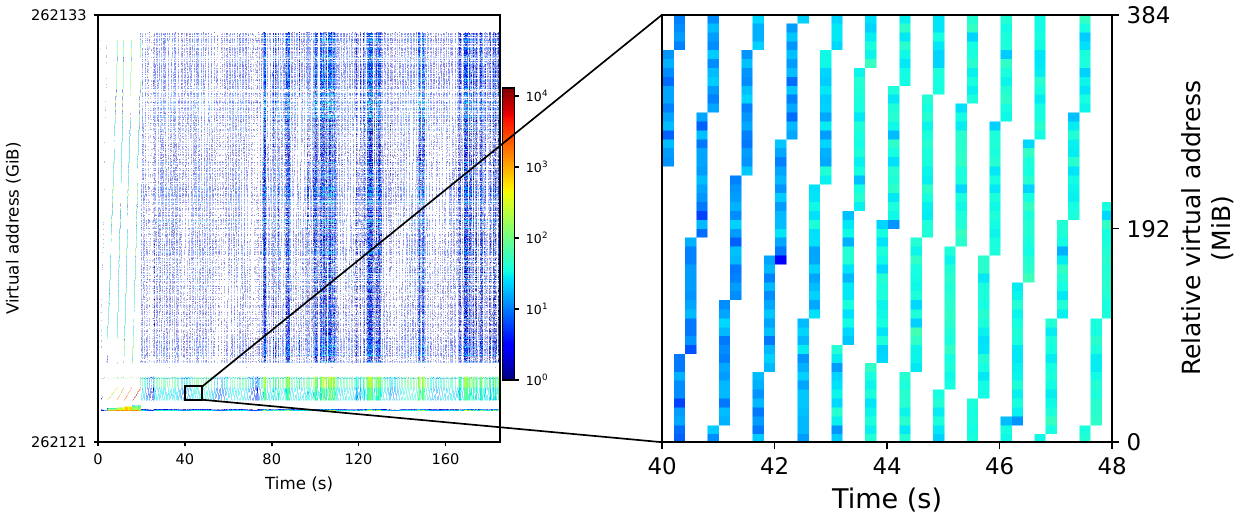}
    \caption{\textcolor{blue}{Qwen3 inference: sampled L1D load-miss overview (left) and KV-cache detail at 40--48~s (right). The detail uses offsets within the 384~MiB KV cache. Both panels use 200~ms by 8~MiB bins and the same logarithmic color scale.}}
    \label{fig:kvcache}
\end{figure*}

\textbf{Pattern III: Block Pattern} represents memory accesses concentrated within specific address regions. In the heatmaps, time-local bursts of memory accesses create rectangular blocks, where contiguous address ranges experience high-frequency access over short periods. Figure~\ref{fig:sparse} also demonstrates block patterns during backward passes. In contrast to the sparse forward pass, backward propagation creates dense rectangular blocks of memory accesses in the activation/gradient region. \textcolor{blue}{This is consistent with tensor-wise operations on gradient buffers during backward computation.}

Block patterns appear in multiple workloads, particularly in LLM and graph processing. 
\textcolor{blue}{Llama inference (Figure~\ref{fig:llama}) exhibits block patterns over broad address ranges, with repeated vertical bands in the upper region after about 40~s.} 
Graph analytics workloads shown in Figures~\ref{fig:tc},~\ref{fig:pr}, and~\ref{fig:cc} also exhibit block patterns corresponding to frequently accessed data structures such as vertex properties or adjacency lists that become hot during iterative computations.

\textbf{\textcolor{blue}{KV Cache Accesses in LLM Inference.}} \textcolor{blue}{To relate strided access patterns to inference computation, we examine KV-cache reuse in Qwen3. Figure~\ref{fig:kvcache} shows its memory access overview and a detailed view of the KV cache. The repeated ascending bands reflect traversal and reuse of cached key and value tensors.}

\subsection{Utilization of Cache and Memory Hierarchy}
\textcolor{blue}{For the eight workloads in Table~\ref{tab:apps}, we further investigate how the memory hierarchy interacts with different memory regions.} These insights are not only useful for optimizing memory-intensive workloads on a multi-tier memory system, but can also guide system design in configuring suitable cache hierarchies for targeted workloads. We focus on three levels: the private L2 cache, shared LLC, and coherence interconnect. We capture hits in each level and correlate them to specific memory regions in each workload. 

\begin{figure}[!htbp]
    \includegraphics[width=0.475\linewidth]{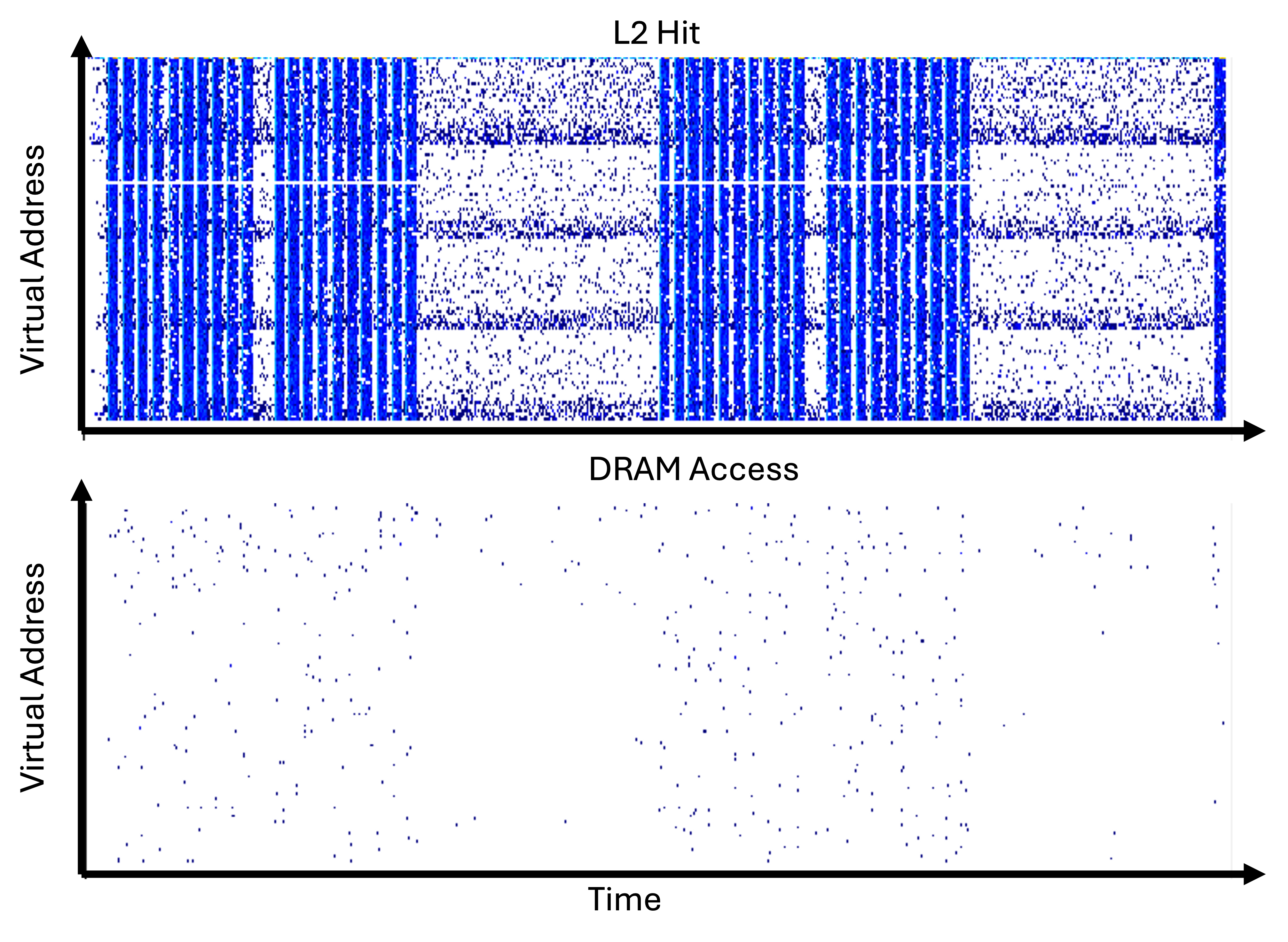}
    \includegraphics[width=0.475\linewidth]{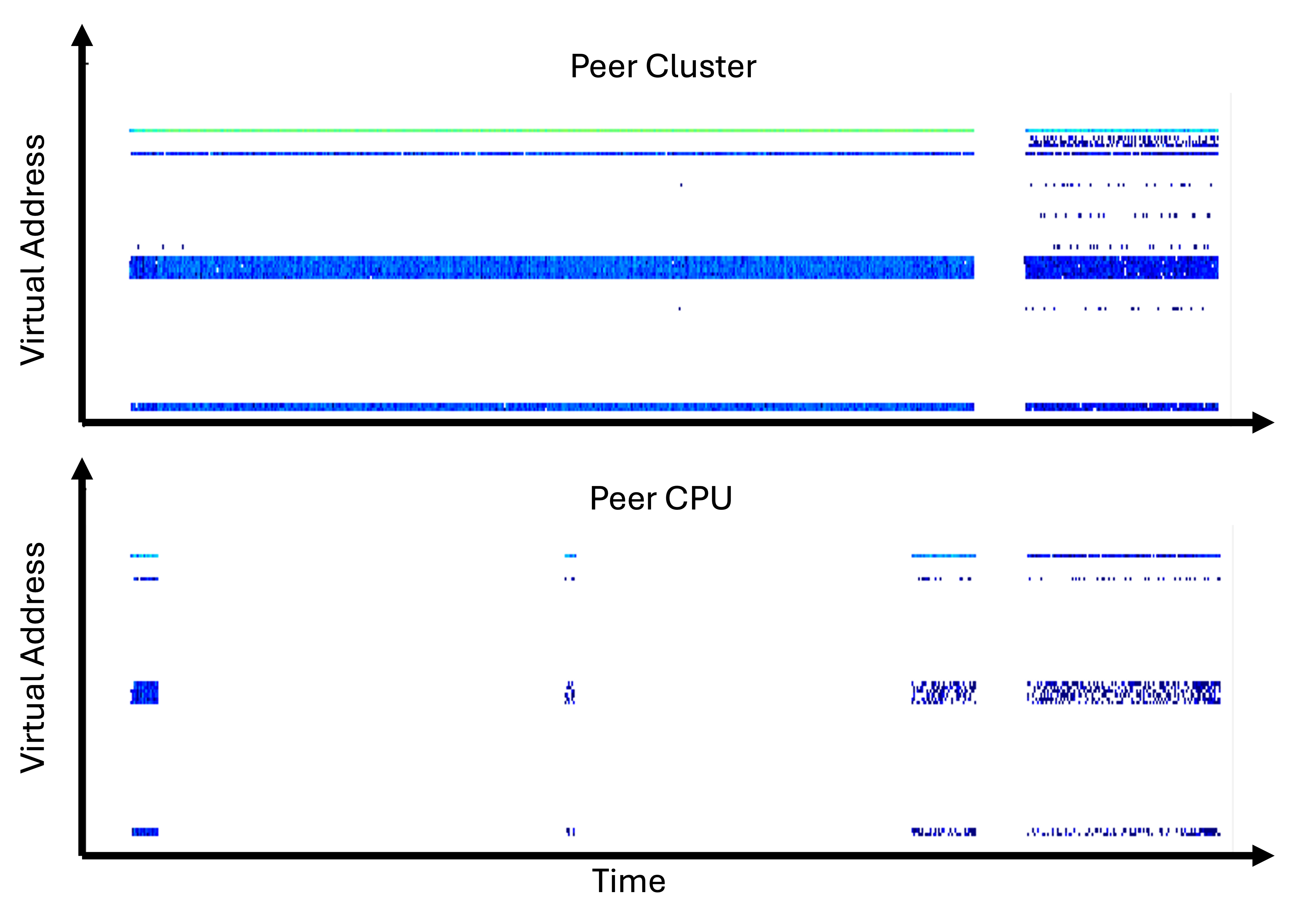}
\caption{Left: L2 hits and DRAM access in LLM GPT2. Right: Peer CPU and Peer Cluster access in Memcached.}
\label{fig:dram-llm}
\end{figure}

The private L2 cache is the most effectively utilized cache level. Figure~\ref{fig:dram-llm} shows the distribution of L2 hits and DRAM access in the same memory region in the LLM GPT2 pretraining workload. By contrasting the two, it is clear that L2 absorbs a vast amount of accesses coming from the middle memory regions, which exhibit an interleaving pattern of dense block access and sparse access as described in Section~\ref{sec:pattern}. \textcolor{blue}{This memory region has very few recorded DRAM-access samples} as shown in the bottom panel of Figure~\ref{fig:dram-llm}, leaving the space almost blank over the execution time. The other workloads also exhibit high utilization of L2 cache. This high utilization of L2 supports the design choice of large (1MB) private L2 for systems targeting such cloud workloads, which is consistent with the initial CloudSuite paper~\cite{ferdman2012clearing}. 

Most workloads can also leverage the shared LLC, except Llama. Llama shows nearly no utilization of LLC, even though it has good utilization of L2, \textcolor{blue}{indicating limited use of the shared LLC in this Llama experiment.} We further investigate memory regions that hit in LLC in each workload and find that mostly only the stride pattern (i.e., horizontal lines in Figure~\ref{fig:pattern}) hits in LLC, while the other patterns, such as \textcolor{blue}{sparse and block patterns}, have relatively lower utilization of it. Note that the LLC on the system is small, i.e., 16 MB compared to \textcolor{blue}{128 MB of aggregate private L2 capacity.} Such a configuration is consistent with the finding that large LLCs are wasted for CloudSuite workloads~\cite{ferdman2012clearing}. 

We analyze data-sharing in these workloads by identifying data fetched from peer CPUs (i.e., cores in the same DSU) and clusters (other DSUs), which indicates the utilization of the coherence interconnect. The right panel of Figure~\ref{fig:dram-llm} presents the same memory regions in Memcached that fetch data from peer DSUs (top) and peer CPU (bottom). Data sharing with other DSUs (peer Cluster) occurs in both phases, while data sharing with peer CPUs only occurs after the initial phase. The initial phase warms the dataset, and as such only consists of set operations that fill up the records, where the second phase consists of a mixture of get and set operations for the data cache. \textcolor{blue}{Llama shows very few samples attributed to Peer CPU or Peer Cluster, indicating limited observed use of these data sources in this experiment.}

\begin{figure}[!htbp]
    \centering
    \includegraphics[width=\linewidth]{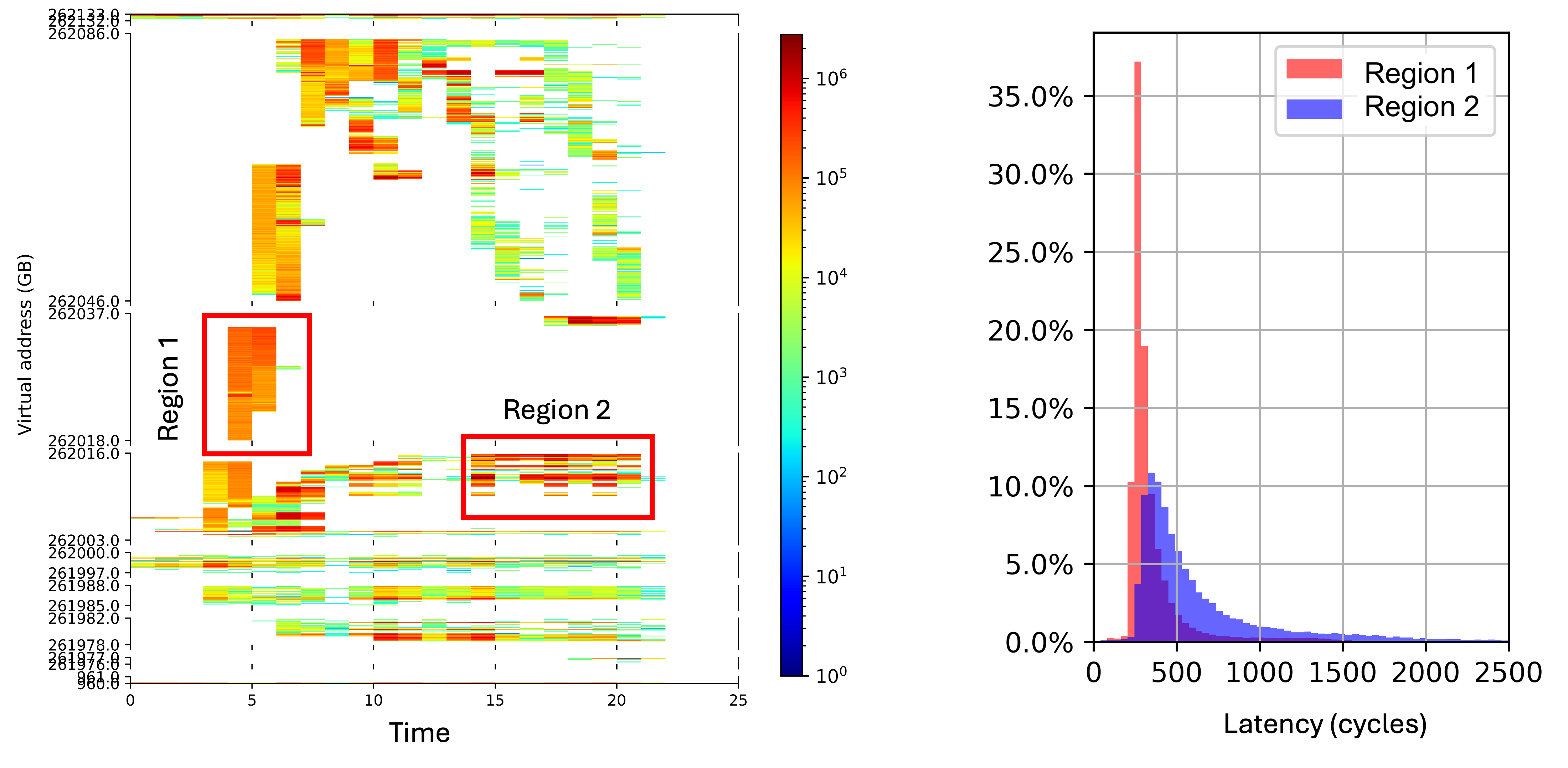}
    \caption{\textcolor{blue}{DRAM-access heatmap and region-specific latency distributions for two hot regions in PageRank.}}
    \label{fig:latency-region}
\end{figure}

\subsection{Memory Region-based Characterization}
\begin{figure*}[!htbp]
    \centering
    \begin{subfigure}{0.24\textwidth}
        \includegraphics[width=\textwidth]{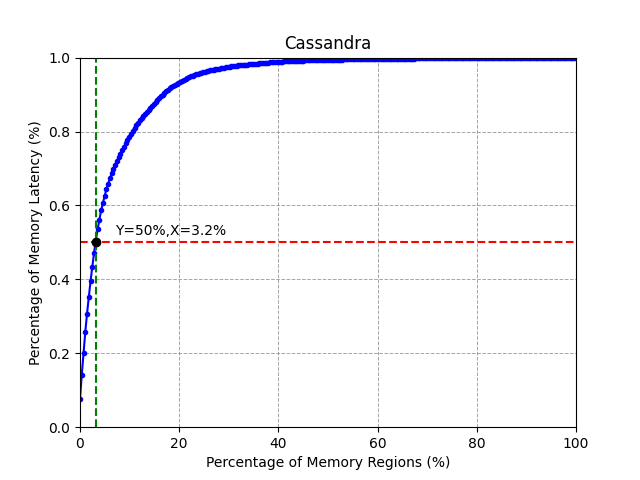}
    \caption{Cassandra}
    \label{fig:lat-cas}
    \end{subfigure}
    \begin{subfigure}{0.24\textwidth}
        \includegraphics[width=\textwidth]{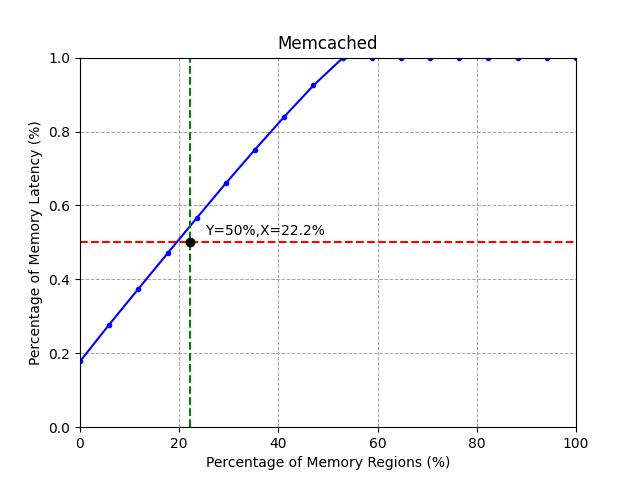}
    \caption{Memcached}
    \label{fig:lat-cached}
    \end{subfigure}    
    \begin{subfigure}{0.24\textwidth}
        \includegraphics[width=\textwidth]{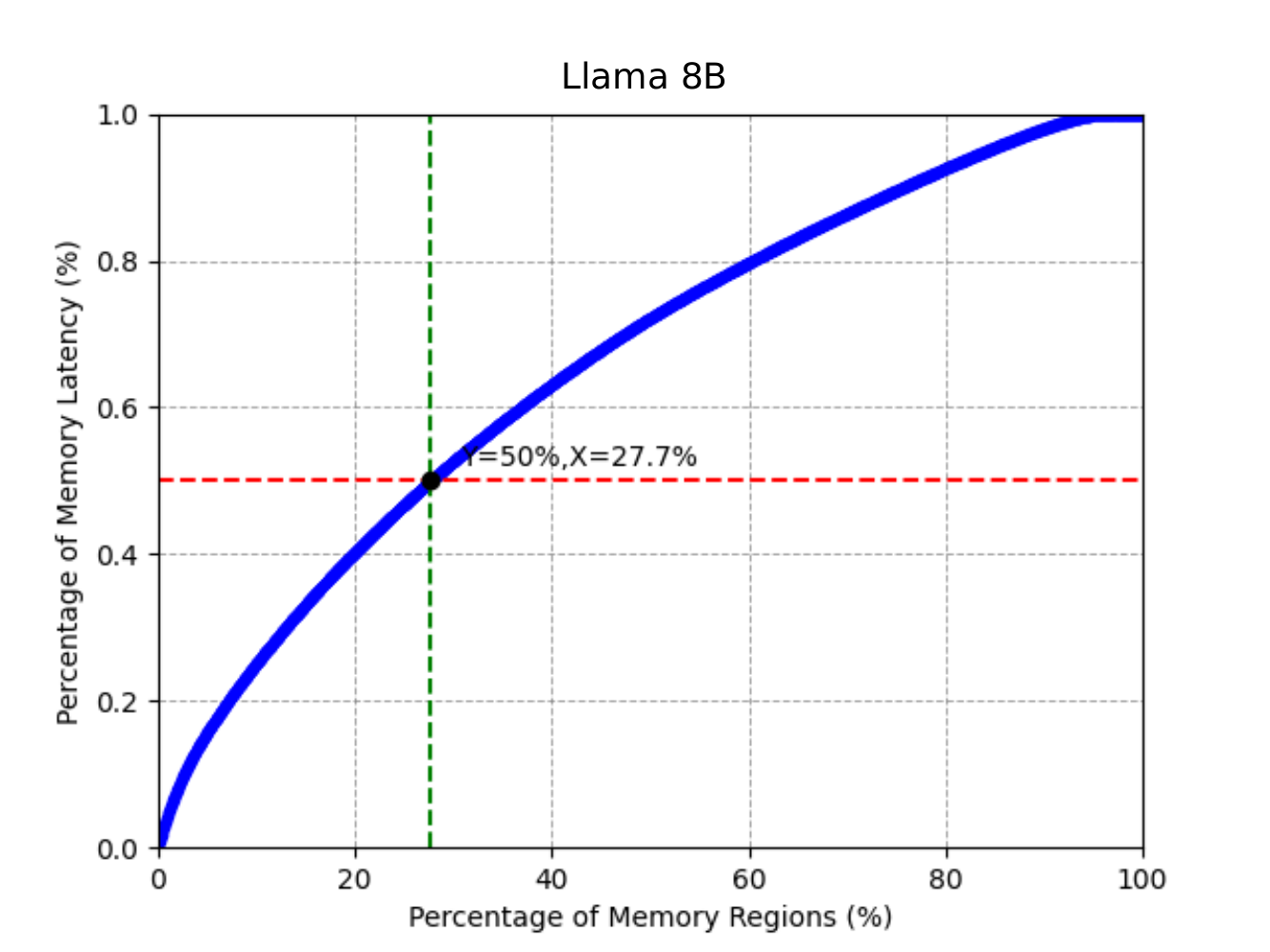}
    \caption{\textcolor{blue}{Llama 8B}}
    \label{fig:lat-llama}
    \end{subfigure}
    \begin{subfigure}{0.24\textwidth}
        \includegraphics[width=\textwidth]{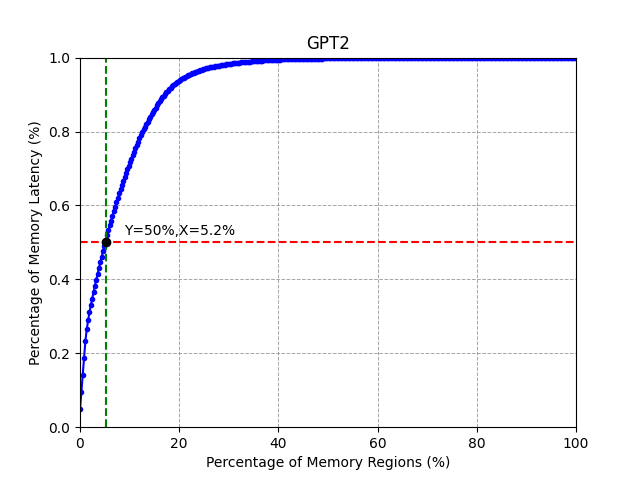}
    \caption{GPT2 Training}
    \label{fig:lat-llm}
    \end{subfigure}    
    \begin{subfigure}{0.24\textwidth}
        \includegraphics[width=\textwidth]{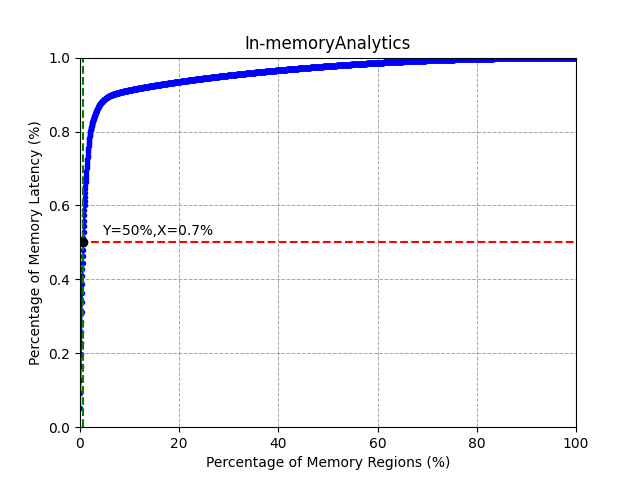}
    \caption{In-Memory Analytics}
    \label{fig:lat-analytics}
    \end{subfigure}
    \centering
    \begin{subfigure}{0.24\textwidth}
        \includegraphics[width=\textwidth]{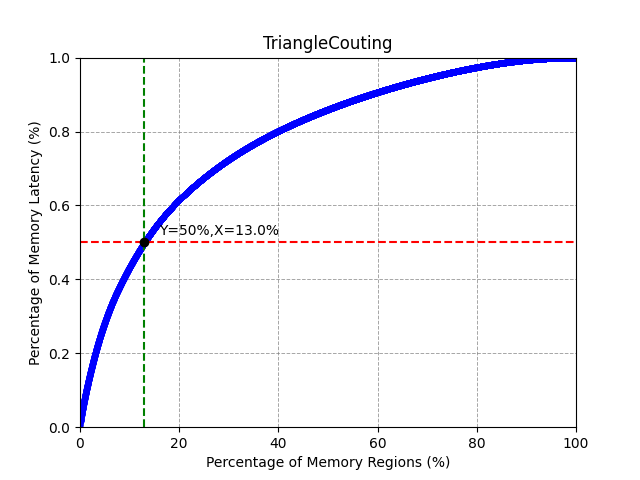}
    \caption{Triangle Counting}
    \label{fig:lat-tr}
    \end{subfigure}
    \begin{subfigure}{0.24\textwidth}
        \includegraphics[width=\textwidth]{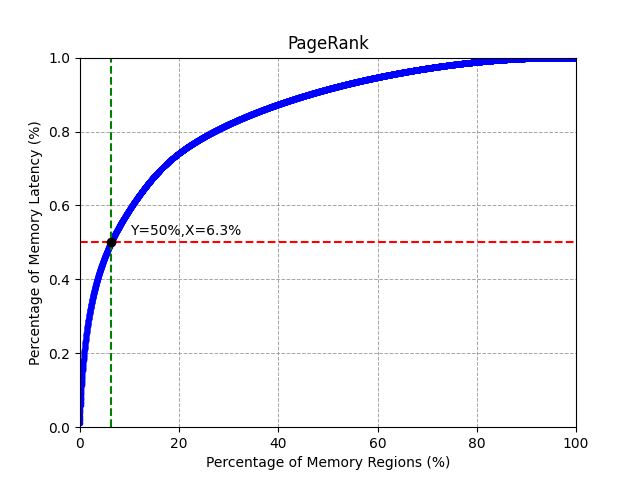}
    \caption{PageRank}
    \label{fig:lat-pr}
    \end{subfigure}
    \begin{subfigure}{0.24\textwidth}
        \includegraphics[width=\textwidth]{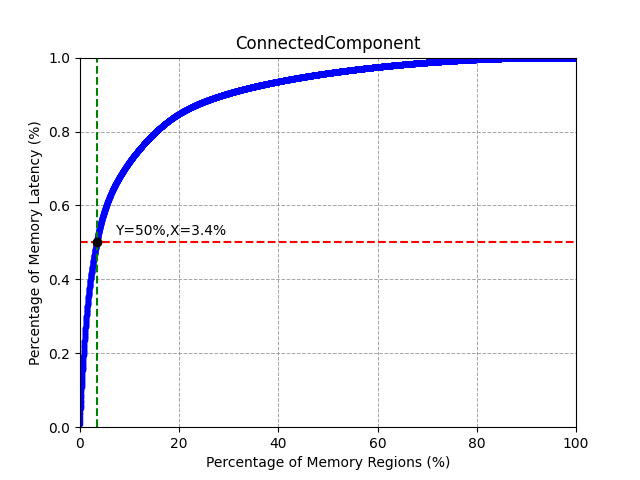}
    \caption{Connected Components}
    \label{fig:lat-cc}
    \end{subfigure}    
    \caption{\textcolor{blue}{Cumulative share of sampled DRAM-access latency, with memory regions ranked by summed latency.}}
    \label{fig:lat-dist}
\end{figure*}

\textcolor{blue}{We next analyze memory regions to identify candidates for placement and migration on multi-tier memory.} A memory region is defined as a contiguous memory space and they are often used as the basis for placement and migration decisions on multi-tier memory. Our first analysis aims to identify the opportunity of running a workload onto multi-tier memory. For this, hot memory regions need to be identified so that they can either be placed into the fast-tier memory statically or be migrated from the slower memory into the fast-tier memory at runtime. A commonly used metric is the count of DRAM accesses to a memory region. One such example is illustrated in the heatmap in Figure~\ref{fig:latency-region}, where hot regions are colored red. 

While access counts provide a good indication of the relative importance among memory regions, our further analysis of region-specific DRAM access latency distribution in two hot regions reveals that they may have very different latency characteristics, as shown in the histogram in Figure~\ref{fig:latency-region}. Region 1, which spans a larger memory size, experiences a short period of hotness, where DRAM latency mostly occurs between 300-500 cycles. In contrast, the latency distribution in Region 2 has a long tail of up to 2500 cycles. \textcolor{blue}{This difference suggests that access pattern and concurrency level can affect the latency observed in different regions.} Since Region 2's size is only one-third of Region 1, it should have higher priority to be considered for the fast memory tier. 

We analyze the latency distribution in all workloads and find that Memcached, Triangle Counting, Page Rank, \textcolor{blue}{and Connected Components} have longer tails in DRAM latency distribution compared to Llama and GPT2. \textcolor{blue}{However, compared to other workloads, Llama shows more samples near 4096 cycles in the recorded latency distribution.}

\subsubsection{Region-Latency Distribution}\label{sec:latency}
We leverage the DRAM latency distribution in memory regions to identify workloads that are promising for multi-tier memory systems. For this, we build the histogram of DRAM accesses in memory regions weighted by their DRAM latency, which better reflects memory access bottlenecks in workloads, compared to the access-only distribution. \textcolor{blue}{Figure~\ref{fig:lat-dist} presents the cumulative share of sampled DRAM-access latency as memory regions are ranked by their summed latency, from highest to lowest. The sum provides a relative ranking of regions; overlapping memory operations mean that it does not measure elapsed stall time.} We identify five workloads showing a highly skewed distribution of DRAM access, including Cassandra, GPT2 training, In-Memory Analytics, page ranking, and connected components. Among them, In-Memory Analytics shows the highest skewed distribution, where $50\%$ latency attributed by merely $0.7\%$ memory regions, as shown in Figure~\ref{fig:lat-analytics}. This workload has nearly all memory regions with the stride access pattern (horizontal lines in Figure~\ref{fig:pattern}) concentrated in small memory regions. With a similar access pattern, Cassandra also has $50\%$ latency attributed to only $3\%$ memory regions. Connected Components and Page Ranking both have their highest latency access associated with memory regions exhibiting stride and block patterns. Finally, the GPT2 training workload has $50\%$ latency attributed to about $5.2\%$ memory regions, which have the stride pattern (streaming). This highly skewed distribution in memory regions is generally favorable on multi-tier memory systems because only a small fraction of memory pages need to be placed in the fast memory tier.

Llama, triangle counting, and Memcached, show a relatively even distribution of DRAM access. \textcolor{blue}{For Llama, about $27.7\%$ of memory regions account for $50\%$ of the sampled latency, as shown in Figure~\ref{fig:lat-llama}. Its latency is therefore less concentrated in a small set of regions than in the highly skewed workloads.} Memcached has most DRAM accesses uniformly distributed to four memory regions associated with the stride pattern. Triangle Counting also has approximately half of its memory regions associated with the block pattern. Therefore, workloads with most memory regions dominated by the block access pattern are less feasible for multi-tier memory than those with a highly skewed distribution. %

\begin{figure*}[!htbp]
    \centering
    \includegraphics[width=0.24\textwidth]{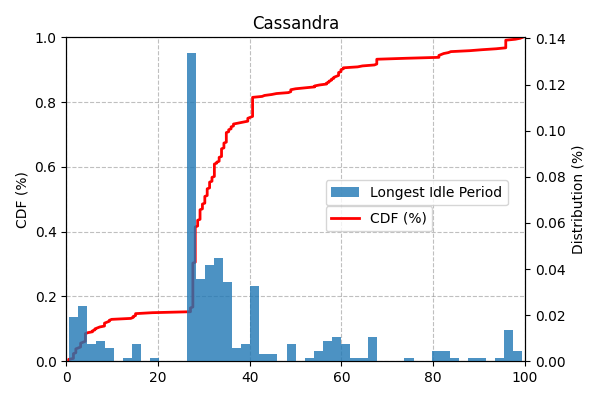}
    \includegraphics[width=0.24\textwidth]{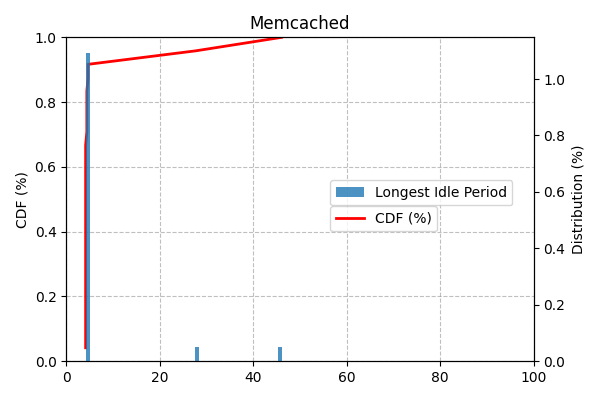}
    \includegraphics[width=0.24\textwidth]{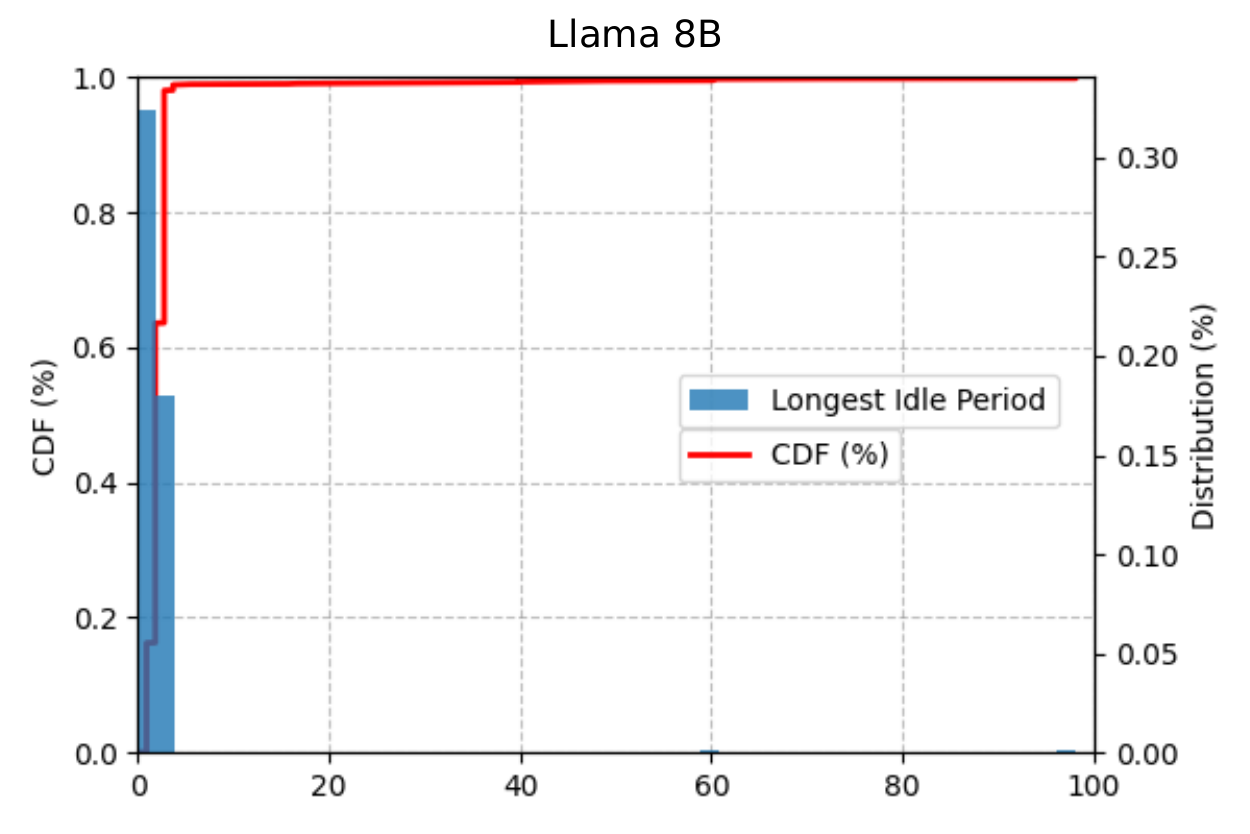}
    \includegraphics[width=0.24\textwidth]{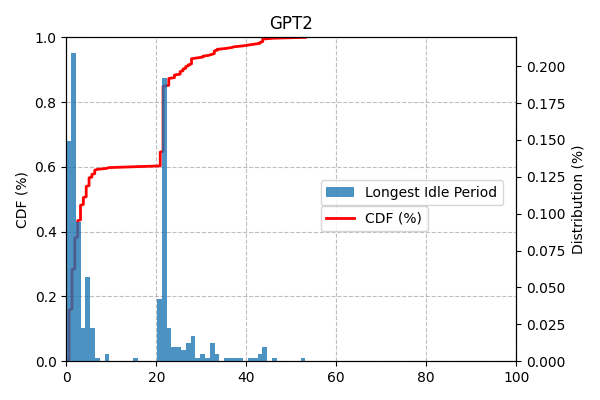}
    \includegraphics[width=0.24\textwidth]{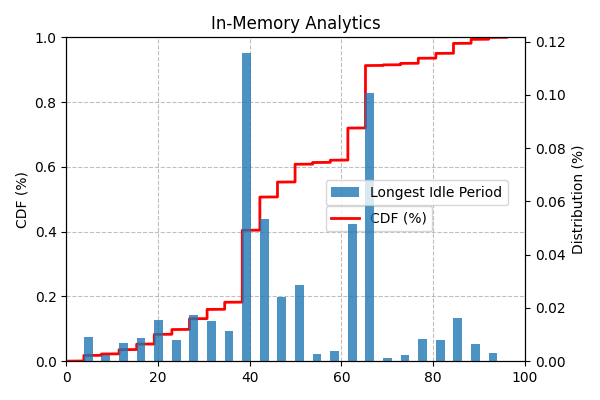}
    \includegraphics[width=0.24\textwidth]{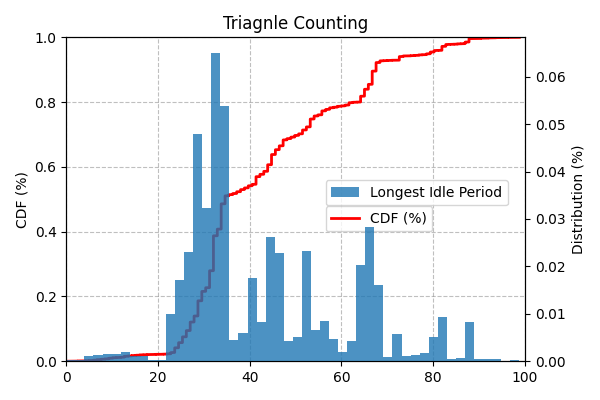}
    \includegraphics[width=0.24\textwidth]{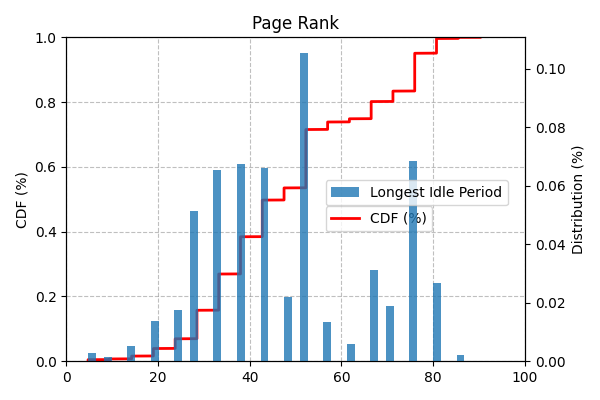}
    \includegraphics[width=0.24\textwidth]{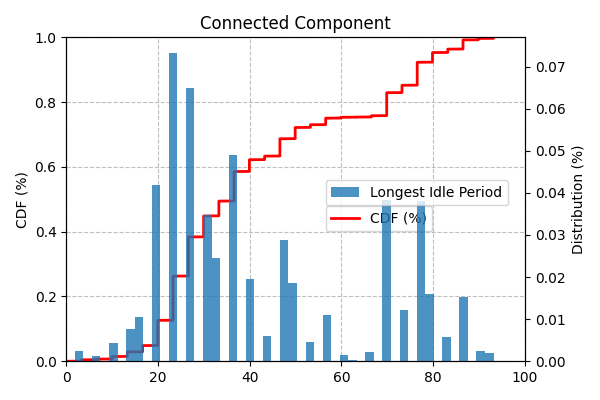}
    \caption{The longest \textcolor{blue}{observed} idle period distribution in memory regions in the \textcolor{blue}{eight workloads}.}
    \label{fig:idle-dist}
\end{figure*}

\subsubsection{Idle Period Distribution}
\label{ref:period}

We propose a new metric\textcolor{blue}{, the \textit{longest idle period fraction},} for assessing the feasibility of data migration on multi-tier memory systems.
The longest idle period fraction is the longest time window \textcolor{blue}{without any recorded DRAM-access samples, normalized by the total execution time.} We compute the metric for each memory region. \textcolor{blue}{This is an observed gap in sampled activity; its duration depends on the sampling period and does not establish that the region received no accesses.}
Note that the longest idle period is different from \textit{the fraction of idle periods}, which sums up all idle periods and cannot distinguish a memory region that frequently interleaves between a busy period (i.e. frequent DRAM accesses) and a short idle period (i.e. \textcolor{blue}{no recorded DRAM-access samples}) and thus, is infeasible for data migration on multi-tier memory systems.

The Longest Idle Period metric can be useful in multiple scenarios. For CXL-enabled multi-tier systems, it can be used to determine whether there are sufficient inactive memory regions that can be offloaded to a slower but less costly memory tier, such as CXL memory pools. It could also be used for assessing whether migrating a memory region is feasible on a system, i.e., if the idle period is long, data migration may overlap with the idle period to mitigate the performance impact on the workload.

We present the distribution of the longest idle period and its CDF in Figure~\ref{fig:idle-dist}. \textcolor{blue}{We use a longest observed idle period above $30\%$ of execution time as a heuristic for identifying migration candidates, although this threshold can be adapted on different target systems.} \textcolor{blue}{From the CDF in Figure~\ref{fig:idle-dist}, Cassandra, In-Memory Analytics, Triangle Counting, Page Rank, and Connected Components have more memory regions with long observed idle periods than Llama and Memcached, although the proportions differ across workloads.} \textcolor{blue}{Llama, instead, has very few memory regions above this threshold, indicating limited opportunities for overlapping migration with long observed idle periods.}  \textcolor{blue}{GPT2 pretraining has longer observed idle periods in some memory regions, consistent with periodically occurring forward-only intervals, which do not need accesses to optimizer states.} Finally, Memcached has less than $5\%$ \textcolor{blue}{of its memory regions above the idle-period threshold.} \textcolor{blue}{Together with the region-latency distributions, these results suggest fewer opportunities for runtime migration in Llama and Memcached, while GPT2 has more concentrated latency and some long observed idle periods.} 

\subsubsection{SPE-guided Optimization on Memory Tiers}
\begin{figure}[!htbp]
    \centering
    \includegraphics[width=\linewidth]{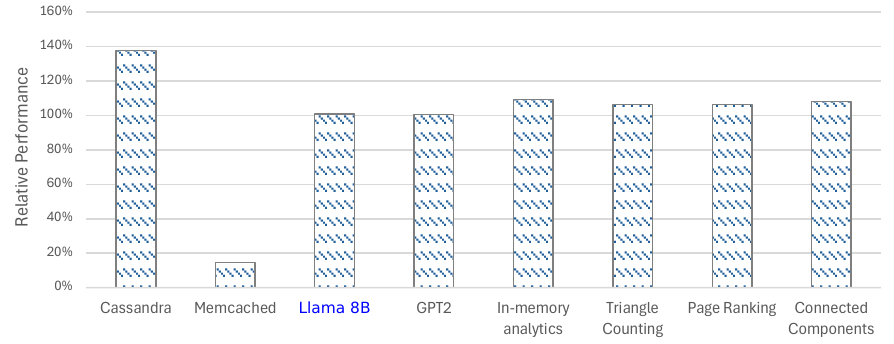}
    \caption{The relative performance of \textcolor{blue}{the eight workloads} with online data migration on multi-tier memory on ARM processor.}
    \label{fig:opt}
\end{figure}
To demonstrate the effectiveness of the proposed memory metrics in optimizing memory usage on ARM processors, we integrated the SPE-based memory profiling feature into a data migration runtime for multi-tier memory. The runtime makes data migration decisions periodically based on memory access samples collected in a time interval. The runtime identifies hot memory regions by selecting the top N memory regions with the most samples from last-level cache misses. Leveraging the physical address of the memory samples, the runtime determines whether a top memory region is already on the fast tier or needs to be migrated from the slow tier. To mitigate the performance interference of memory region migration to the running program, the runtime limits the maximum data size that can be migrated in a time interval. In this experiment, we use \verb|move_pages| system call to migrate all selected memory regions in one call and set the migration frequency to be every second. \textcolor{blue}{For this migration experiment, we used a two-NUMA ARMv8 processor based testbed, on which} we emulated a two-tier memory system by injecting memory traffic to one NUMA node and limiting the free memory space on the other NUMA node. Effectively, the emulated multi-tier memory has a memory bandwidth ratio of $1.8\times$ between the fast and the slower tier, and a memory capacity ratio of $2\times$. 

The results of SPE-guided optimization on multi-tier memory are presented in Figure~\ref{fig:opt}. The memory analysis of region latency distribution in Section~\ref{sec:latency} identifies workloads suitable for multi-tier memory systems as those with skewed distribution. Consistent with this memory analysis, performance improvement is observed in the suitable workloads such as page ranking and connected components. In contrast, \textcolor{blue}{Llama and Memcached are less suitable for the tested migration policy,} and they achieve no performance benefit. \textcolor{blue}{The longest observed idle period metric identifies workloads with few long gaps in sampled activity, for which runtime migration may cause interference.} For instance, Memcached is identified as \textcolor{blue}{one of the least promising workloads} for runtime migration because it has very few \textcolor{blue}{regions with long observed idle periods}. Consequently, in Figure~\ref{fig:opt}, Memcached suffers from a huge performance loss, as is quantified with its built-in 99th percentile latency, from runtime migration. \textcolor{blue}{Moreover, Cassandra has a substantial proportion of memory regions with long observed idle periods, making it a candidate for runtime migration.} The results confirm that Cassandra achieves the highest performance improvement. In summary, the results show the usefulness of the memory region based metrics as introduced in \textcolor{blue}{Sections~\ref{sec:latency} and~\ref{ref:period}} in guiding optimization on multi-tier memory.

\section{Related Works}
\label{sec:related}

Existing work explores memory access profiling with OS-based and PMU-based method. \textcolor{blue}{OS-based methods observe page-level access information, providing limited visibility into individual memory operations.} Page poisoning techniques~\cite{asplos17_thermostat, badgertrap, qi2025chrono} track specific pages through customized fault handlers but incur substantial overhead from fault handling. 
Other work explores PTE-scan based profiling methods~\cite{park2019profiling, nair2024telescope, maruf2023tpp, eurosys24-mtm, ipdps21:dancing}, which clear the PTE access bit and periodically scan the page table to determine if any page has been accessed between two scans. However, such method cannot detect repeated page accesses within profiling intervals, missing true hot regions.
Methods using OS structures like active/inactive queues~\cite{Yan:2019:NPM:3297858.3304024, atc21_autotiering} are lightweight but provide limited information.
In contrast, our ARM SPE-based approach provides accurate, fine-grained memory access profiling with \textcolor{blue}{supported} memory hierarchy information (L1, L2, and TLB level accesses). \textcolor{blue}{As demonstrated in our evaluation, we achieve higher hot-region precision and lower overhead than DAMON under the tested configurations, while recall depends on the workload.}


PMU based techniques rely on the performance monitoring units (PMUs) in modern processors, including Intel PEBS, AMD IBS, and Arm SPE. Among these, PEBS-based profiling methods have been widely used in NUMA bottleneck analysis~\cite{liu2015scaanalyzer}, page hotness identification~\cite{maruf2023tpp}, and memory traces~\cite{boehme2016caliper}. Studies show PEBS offers accurate event sampling while IBS provides richer per-sample data~\cite{sasongko2023precise}, though sampling bias and overhead remain challenges~\cite{xu2019can, yi2020precision, soramichi2017quantitative}. \textcolor{blue}{ARM SPE combines supported cache and TLB events, data-source information, and operation latency in a sample record, enabling memory characterization on ARM platforms.}

\section{Conclusions}
In this work, we present a high-resolution memory characterization methodology based on the ARM SPE feature, addressing the increasing deployment of ARM-based processors in data centers. Motivated by new capabilities enabled by memory-centric sample tracing, we propose a set of new metrics, such as latency profiles and longest idle periods, for assessing the opportunity and feasibility of a workload on multi-tier memory systems. We provide a validated high-performance, low-overhead implementation of the characterization method. By evaluating eight memory-intensive workloads, including CloudSuite and LLM applications, on an ARMv8-based platform, our results highlight the usefulness of the proposed approach in studying a cloud-oriented cache hierarchy and in guiding the memory optimizations.

\begin{acks}
This research is supported by the European Commission under the Horizon project OpenCUBE (101092984).
\end{acks}

\bibliographystyle{ACM-Reference-Format}
\bibliography{references}

\end{document}